\documentclass{emulateapj}

\shorttitle{Uncertainties in the $\nu$p-process}

\shortauthors{Wanajo et al.}

\begin{document}

\title{Uncertainties in the $\nu$\lowercase{p}-process: supernova dynamics versus
nuclear physics}

\author{Shinya Wanajo\altaffilmark{1, 2},
        Hans-Thomas Janka\altaffilmark{2},
        and 
        Shigeru Kubono\altaffilmark{3}
        }

\altaffiltext{1}{Technische Universit\"at M\"unchen,
        Excellence Cluster Universe,
        Boltzmannstr. 2, D-85748 Garching, Germany;
        shinya.wanajo@universe-cluster.de}

\altaffiltext{2}{Max-Planck-Institut f\"ur Astrophysik,
        Karl-Schwarzschild-Str. 1, D-85748 Garching, Germany;
        thj@mpa-garching.mpg.de}

\altaffiltext{3}{Center for Nuclear Study, University of Tokyo,
        RIKEN Campus, 2-1 Hirosawa, Wako, Saitama 351-0198, Japan;
        kubono@cns.s.u-tokyo.ac.jp}


\begin{abstract}
  We examine how the uncertainties involved in supernova dynamics as
  well as in nuclear data inputs affect the $\nu$p-process in the
  neutrino-driven winds. For the supernova dynamics, we find that the
  wind termination by the preceding dense ejecta shell, as well as the
  electron fraction ($Y_\mathrm{e, 3}$; at $3\times 10^9$~K) play a
  crucial role. A wind termination within the temperature range of
  $(1.5-3) \times 10^9$~K greatly enhances the efficiency of the $\nu
  $p-process. This implies that the early wind phase, when the
  innermost layer of the preceding supernova ejecta is still $\sim
  200-1000$~km from the center, is most relevant to the
  $\nu$p-process. The outflows with $Y_\mathrm{e, 3} = 0.52-0.60$
  result in the production of the p-nuclei up to $A=108$ with
  interesting amounts. Furthermore, the p-nuclei up to $A=152$ can be
  produced if $Y_\mathrm{e, 3} = 0.65$ is achieved. For the nuclear
  data inputs, we test the sensitivity to the rates relevant to the
  breakout from the pp-chain region ($A < 12$), to the $(n, p)$ rates
  on heavy nuclei, and to the nuclear masses along the $\nu$p-process
  pathway. We find that a small variation of the rates of
  triple-$\alpha$ and of the $(n, p)$ reaction on $^{56}$Ni leads to a
  substantial change in the p-nuclei production. We also find that
  $^{96}$Pd ($N=50$) on the $\nu$p-process path plays a role as a
  second seed nucleus for the production of heavier p-nuclei. The
  uncertainty in the nuclear mass of $^{82}$Zr can lead to a factor of
  two reduction in the abundance of the p-isotope $^{84}$Sr.
\end{abstract}


\section{Introduction}
The astrophysical origin of the proton-rich isotopes of heavy elements
(p-nuclei) is not fully understood. The most successful model to date,
the photo-dissociation of pre-existing neutron-rich isotopes
($\gamma$-process) in the oxygen-neon layer of core-collapse
supernovae (or in their pre-collapse stages), cannot explain the
production of some light p-nuclei including $^{92, 94}$Mo and $^{96,
  98}$Ru \citep{Woos1978, Pran1990, Raye1995, Raus2002, Haya2008}. The
recent discovery of a new nucleosynthetic process, the $\nu
$p-process, has dramatically changed this difficult situation
\citep{Froe2006, Froe2006b, Prue2006, Wana2006}. In the early
neutrino-driven winds of core-collapse supernovae, $\bar{\nu}_e$
capture on free protons gives rise to a tiny amount of free neutrons
in the proton-rich matter. These neutrons induce the $(n, p)$
reactions on the $\beta^+$-waiting point nuclei along the classical
rp-process path ($^{64}$Ge, $^{68}$Se, and $^{72}$Kr), which bypass
these nuclei (with the $\beta^+$-decay half-lives of 1.06~min, 35.5~s,
and 17.1~s, respectively). \citet{Wana2006} has shown that the
p-nuclei up to $A\sim 110$, including $^{92, 94}$Mo and $^{96, 98}$Ru,
can be produced by the $\nu$p-process in the neutrino-driven winds
within reasonable ranges of the model parameters.

All the recent hydrodynamic studies of core-collapse supernovae with
neutrino transport taken into account suggest that the bulk of early
supernova ejecta is proton rich \citep{Jank2003, Lieb2003, Bura2006,
  Kita2006, Fisc2010, Hued2010}. This supports the $\nu$p-process
taking place in the neutrino-driven winds of core-collapse
supernovae. However, different works end up with somewhat different
outcomes. \citet{Froe2006} showed that the p-nuclei up to $A\sim 80$
were produced with the one-dimensional, \textit{artificially} induced
explosion model of a $20 M_\odot$ star, while \citet{Prue2006}
obtained up to $A\sim 100$ with the two-dimensional,
\textit{artificially} induced explosion model of a $15 M_\odot$
star. On the contrary, \citet[][also S. Wanajo et al., in
preparation]{Wana2009} found negligible contribution of the
$\nu$p-process to the production of p-nuclei with the one-dimensional,
\textit{self-consistently} exploding model of a $9 M_\odot$ star
\citep[electron-capture supernova,][]{Kita2006, Hued2010}. These
diverse outcomes indicate that the $\nu$p-process is highly sensitive
to the physical conditions of neutrino-driven winds.

Besides the supernova conditions, there could be also uncertainties in
some key nuclear rates, in particular of $(n, p)$ reactions, because
no attention was paid to neutron capture reactions on proton-rich
nuclei before the discovery of the $\nu $p-process. Uncertainties in
some reactions relevant to the breakout from the pp-chain region ($A <
12$), which affect the proton-to-seed ratio at the onset of
$\nu$p-processing, might also influence the nucleosynthetic
outcomes. There are still a number of isotopes without experimental
nuclear masses on the $\nu$p-process pathway \citep{Webe2008}.

Our goal in this paper is to examine how the variations of supernova
conditions as well as of nuclear data inputs influence the global
trend of the $\nu$p-process. The paper is organized as follows. In
\S~2, a basic picture of the $\nu$p-process is outlined. A
semi-analytic neutrino-driven wind model and an up-to-date reaction
network code are described, which are used in this study (\S~3). We
take the wind-termination radius (or temperature), the neutrino
luminosity, the neutron-star mass, and the electron fraction as the
key parameters of supernova conditions (\S~4). In previous studies
\citep{Froe2006, Froe2006b, Prue2006, Wana2006}, some of these
parameters were varied to test their sensitivities, but only for
limited cases. In particular, the effect of wind termination has not
been discussed at all in previous studies. As the key nuclear
reactions, we take triple-$\alpha$, $^7$Be$(\alpha, \gamma)^{11}$C,
$^{10}$B$(\alpha, p)^{13}$C (all relevant to the breakout from the
pp-chain region), and the $(n, p)$ reactions on $^{56}$Ni, $^{60}$Zn,
and $^{64}$Ge (\S~5). Sensitivities of the masses of the nuclei along
the $\nu$p-process path are also discussed. We then discuss the
possible role of the $\nu$p-process as the astrophysical origin of the
p-nuclei (\S~6). A summary of our results follows in \S~7.

\section{Basic Picture of the $\nu$\lowercase{p}-Process}

The ``$\nu$p-process'' was first identified in \citet{Froe2006}, and
the term was introduced by \citet{Froe2006b} and is synonymous with
the ``neutrino-induced rp-process'' in the subsequent works
\citep{Prue2006, Wana2006}. This is a similar process to the
\textit{classical} rp-process first proposed by \citet{Wall1981}. The
$\nu$p-process is, however, essentially a new nucleosynthetic process
exhibiting a number of different aspects compared to the classical
rp-process. The $\nu$p-process starts with the seed nucleus $^{56}$Ni
(\textit{not} $^{64}$Ge, the first $\beta^+$-waiting-point nucleus in
the classical rp-process pathway), assembled from free nucleons in
nuclear quasi-equilibrium (QSE) during the initial high temperature
phase ($T_9 > 4$; where $T_9$ is the temperature in units of
$10^9$~K). The $\nu$p-process is therefore a \textit{primary} process,
which needs no pre-existing seeds. When the temperature decreases
below $T_9 = 3$ (defined as the onset of a $\nu$p-process in this
study) and QSE freezes out, the $\nu$p-process starts.

Neutrino capture on free protons, $p(\bar{\nu}_\mathrm{e}, e^+)n$, in
a proton-rich neutrino-driven wind gives rise to a tiny amount of free
neutrons ($10^{-11}-10^{-12}$ in mass fraction). These neutrons
immediately induce the exchange reaction, $(n, p)$, and in part
radiative neutron capture, $(n, \gamma)$, on the seed nucleus
$^{56}$Ni and subsequent heavier nuclei with decay timescales of a few
ms, well below the expansion timescale of the wind and well below the
$\beta^+$-decay lifetimes of these nuclei. The nuclear flow proceeds
with combination of radiative proton captures, $(p, \gamma)$, and
neutron captures, the latter replacing the role of $\beta^+$-decays in
the classical rp-process.

A large number of free protons relative to that of $^{56}$Ni at $T_9 =
3$, which allows for neutron capture on the seed nuclei, is required
to initiate the $\nu$p-process. High entropy and short expansion
timescale of the ejecta make the triple-$\alpha$ process, bridging
from light ($A < 12$) to heavy ($A \ge 12$) nuclei, less effective and
help to leave a large number of free protons at the onset of the
$\nu$p-process. It should be noted, however, that proton-rich matter
freezing out from nuclear statistical equilibrium (NSE) mainly
consists of $^{56}$Ni and \textit{free protons} \citep{Seit2008}. This
is a fundamental difference from a (moderately) neutron-rich NSE,
where no free neutrons exist at freezeout. This makes the requirements
for entropy and expansion timescale less crucial, compared to the case
of r-process, allowing for the $\nu$p-process taking place with
typical wind conditions \citep{Froe2006, Prue2006, Wana2006}.

Unlike the r-process, the $\nu$p-process is not terminated by the
exhaustion of free protons, but by the temperature decrease below $T_9
= 1.5$ (defined as the end of a $\nu$p-process), where proton capture
slows due to the Coulomb barrier. The end of $\nu$p-processing is thus
a proton-rich freezeout. For this reason, the proton-to-seed ratio,
$Y_\mathrm{p}/Y_\mathrm{h}$ (the number per nucleon for free protons
divided by that for nuclei with $Z > 2$) at $T_9 = 3$ does not
necessary serve as a useful guide for the strength of the
$\nu$p-process as the neutron-to-seed ratios are in the case of the
r-process. Rather, the number ratio of free neutrons created by
$p(\bar{\nu}_\mathrm{e}, e^+)n$ (for $T_9 \le 3$) relative to the seed
nuclei (at $T_9 = 3$), $\Delta_\mathrm{n}$, can be a useful (but still
crude) measure for the $\nu$p-process as proposed by
\citet{Prue2006}. Note that each neutron capture by $(n, p)$ is
immediately followed by one or two radiative proton captures,
increasing the atomic masses by one or two units. Similar to eq.~(2)
in \citet{Prue2006}, we define
\begin{eqnarray}
\Delta_\mathrm{n}
= \frac{Y_\mathrm{p} n_{\bar{\nu}_\mathrm{e}}}{Y_\mathrm{h}},
\end{eqnarray}
where $Y_\mathrm{p}$ (equal to the mass fraction of free protons,
$X_\mathrm{p}$) and $Y_\mathrm{h}$ are the values at $T_9 = 3$. The
net number of $\bar{\nu}_\mathrm{e}$ captured per free proton for $T_9
\le 3$, $n_{\bar{\nu}_\mathrm{e}}$, is defined as
\begin{eqnarray}
n_{\bar{\nu}_\mathrm{e}}
= \int_{T_9 \le 3} \lambda_{\bar{\nu}_\mathrm{e}} \, dt,
\end{eqnarray}
where $\lambda_{\bar{\nu}_\mathrm{e}}$ is the rate for
$p(\bar{\nu}_\mathrm{e}, e^+)n$. The seed, a double magic nucleus
$^{56}$Ni, remains the most abundant heavy nucleus throughout the
$\nu$p-process. Therefore, only a fraction of $^{56}$Ni is consumed
for the production of heavier nuclei. For this reason, $\Delta_n \sim
10$ is enough for the production of nuclei with $A \sim 100-110$, as
we will see in the subsequent sections.

The $\nu$p-process flow passes through the even-even nuclei up to $Z =
N \sim 40$ and gradually deviates toward the $Z < N$ region.
As the flow proceeds toward higher $Z$ nuclei, and as the temperature
decreases, $(n, \gamma)$ competes with $(n, p)$, owing to the latter
having a Coulomb barrier in its exit channel. When $\Delta_\mathrm{n}$
is large enough,
the flow eventually approaches the $\beta$-stability line, and even
crosses into the neutron-rich region. The latter happens when the net
number of $\bar{\nu}_\mathrm{e}$ captured per free proton after the
$\nu$p-process, defined as
\begin{eqnarray}
n_{\bar{\nu}_\mathrm{e}}'
= \int_{T_9 \le 1.5} \lambda_{\bar{\nu}_\mathrm{e}} \, dt,
\end{eqnarray}
is not negligible compared to $n_{\bar{\nu}_\mathrm{e}}$. The end
point of the $\nu$p-process is thus determined by the supernova
dynamics, which enters into Eq.~(3) through
$\lambda_{\bar{\nu}_\mathrm{e}} \propto r^{-2}$ ($r$ is the radius
from the center), rather than by the nature of nuclear physics as in
case of the classical rp-process.




\section{Neutrino-Driven Wind Model and Reaction Network}

\begin{figure}
\epsscale{1.0}
\plotone{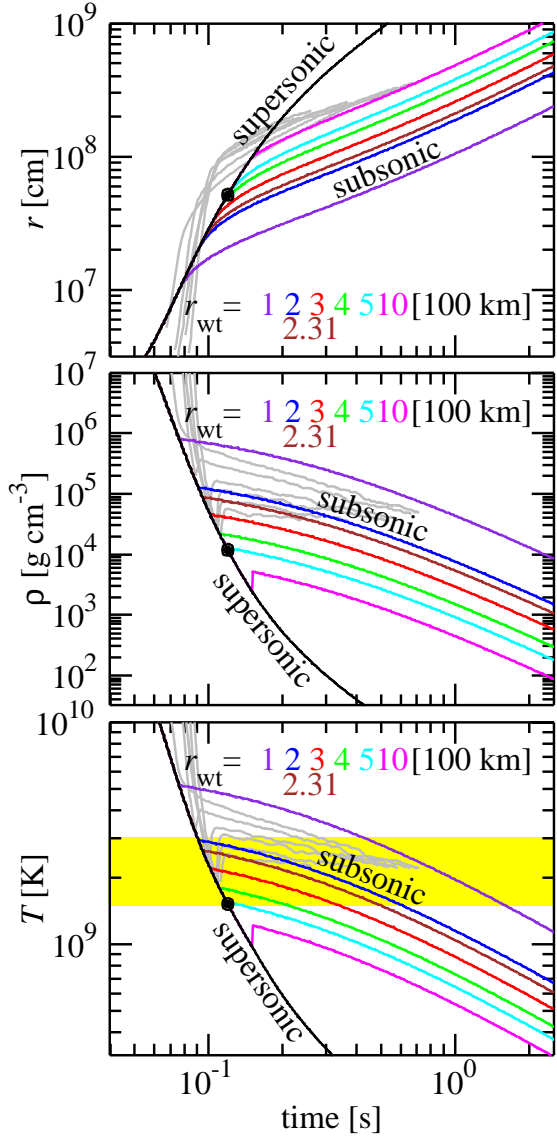}
\caption{Radius (top), density (middle), and temperature (bottom) as a
  function of time (set to 0 at the neutrino sphere) for
  $M_\mathrm{ns} = 1.4\, M_\odot$ and $L_\nu = 1\times 10^{52}\,
  \mathrm{erg\ s}^{-1}$. Subsonic outflows after wind termination at
  $r_\mathrm{wt} = 100, 200, 231, 300, 400, 500$, and 1000~km are
  color coded. The black line shows the supersonic outflow without
  wind termination. In each panel, a filled circle marks the sonic
  point. The yellow band in the bottom panel indicates the temperature
  range ($T_9 = 1.5 - 3$) relevant to the $\nu$p-process. The wind
  trajectories from hydrodynamical results by \citet[][gray
  lines]{Bura2006}, used in \citet{Prue2006}, are compared with our
  models.}
\end{figure}

The thermodynamic trajectories of neutrino-driven outflows are obtained
using a semi-analytic, spherically symmetric, general relativistic model
of neutrino-driven winds. This model has been developed in previous
r-process \citep{Wana2001,Wana2007} and $\nu$p-process \citep{Wana2006}
studies. Here, we describe several modifications added to the previous
version.

The equation of state for ions (ideal gas) and arbitrarily degenerate,
arbitrarily relativistic electrons and positrons is taken from
\citet{Timm2000}. The root-mean-square averaged energies of neutrinos
are taken to be 12, 14, and 14~MeV, for electron, anti-electron, and
the other types of neutrinos, respectively, in light of a recent
self-consistently exploding model of a $9 M_\odot$ star
\citep{Kita2006, Hued2010, Muel2010}. These values are consistent with
other recent studies for more massive progenitors \citep{Fisc2010},
but substantially smaller than those taken in previous works
\citep[e.g., 12, 22, and 34~MeV in ][]{Wana2001}. The mass ejection
rate $\dot M$ at the neutrino sphere is determined such that the
outflow becomes supersonic (i.e., \textit{wind}) through the sonic
point.

The neutron star mass $M_\mathrm{ns}$ is taken to be $1.4\, M_\odot$
for our standard model. The radius of the neutrino sphere is assumed
to be $R_\nu (L_\nu) = (R_{\nu 0} - R_{\nu 1}) (L_\nu/L_{\nu 0}) +
R_{\nu 1}$ as a function of the neutrino luminosity $L_\nu$ (taken to
be the same for all the flavors), where $R_{\nu 0} = 30\,
\mathrm{km}$, $R_{\nu 1} = 10\, \mathrm{km}$, and $L_{\nu 0} =
10^{52.6} = 3.98 \times 10^{52} \, \mathrm{ergs\ s}^{-1}$. This
roughly mimics the evolution of $R_\nu$ in recent hydrodynamic
simulations \citep[e.g.,][]{Bura2006, Arco2007}. The wind solution is
obtained with $L_\nu = 1 \times 10^{52}\, \mathrm{erg\ s}^{-1}$
($R_\nu = 12.5$~km) for the standard model. The time variations of
radius $r$ from the center, density $\rho$, and temperature $T$ for
the standard model are shown in Figure~1 (black line).

The time variations of $r$, $\rho$, and $T$ after the wind-termination
by the preceding supernova ejecta are calculated as follows. This phase
is governed by the evolution of the preceding slowly outgoing ejecta,
independent of the wind solution. In light of recent hydrodynamical
calculations \citep[e.g.,][]{Arco2007}, we assume the time evolution of
the outgoing ejecta to be $\rho \propto t^{-2}$ and $T \propto
t^{-2/3}$, where $t$ is the post-bounce time. With these relations, we
have
\begin{eqnarray}
\rho(t)
& = & \rho_\mathrm{wt}\left(\frac{t}{t_\mathrm{wt}}\right)^{-2},\\
T(t)
& = & T_\mathrm{wt}\left(\frac{t}{t_\mathrm{wt}}\right)^{-\frac{2}{3}},\\
r(t)
& = & r_\mathrm{wt}\left[
  1 - \frac{u_\mathrm{wt}t_\mathrm{wt}}{r_\mathrm{wt}}
  + \frac{u_\mathrm{wt}t_\mathrm{wt}}{r_\mathrm{wt}}
  \left(\frac{t}{t_\mathrm{wt}}\right)^3
  \right]^\frac{1}{3},\\
u(t)
& = & u_\mathrm{wt}\left[
  1 - \frac{u_\mathrm{wt}t_\mathrm{wt}}{r_\mathrm{wt}}
  + \frac{u_\mathrm{wt}t_\mathrm{wt}}{r_\mathrm{wt}}
  \left(\frac{t}{t_\mathrm{wt}}\right)^3
  \right]^{-\frac{2}{3}}\left(\frac{t}{t_\mathrm{wt}}\right)^2,
\end{eqnarray}
for $t > t_\mathrm{wt}$, where $t_\mathrm{wt}$, $u_\mathrm{wt}$,
$r_\mathrm{wt}$, $\rho_\mathrm{wt}$, and $T_\mathrm{wt}$ are the time,
velocity, radius, density, and temperature, respectively, just after the
wind-termination. Equation~(7) represents the time variation of velocity
after the wind-termination. In case that $r_\mathrm{wt}$ is larger than
that at the sonic point, $r_\mathrm{s}$, the Rankine-Hugoniot shock-jump
conditions are applied at $r_\mathrm{wt}$ to obtain $u_\mathrm{wt}$,
$\rho_\mathrm{wt}$, and $T_\mathrm{wt}$ \citep[see,
e.g.,][]{Arco2007,Kuro2008}. Equations~(6) and (7) are obtained from
equation~(4) with the steady-state condition, i.e., $r^2 \rho u =$
constant \citep[see][]{Pano2009}. Note that equations~(6) and (7) gives
$r(t) \propto t$ and $u(t) = \mathrm{constant}$ for $t \gg
t_\mathrm{wt}$. In order to obtain $t$ in equations~(4)-(7) for a given
trajectory with $L_\nu$, the time evolution of $L_\nu$ at the neutrino
sphere is assumed to be $[L_\nu(t)]_{r = R_\nu} = L_{\nu 0}
(t/t_0)^{-1}$, where $t > t_0 = 0.2$~s \citep{Wana2006}. With this
relation, the post-bounce time is determined to be $t = (L_{\nu
0}/L_\nu) t_0 + t_\mathrm{loc}$, where $t_\mathrm{loc}$ is the local
time in each wind trajectory ($t_\mathrm{loc} = 0$ at the neutrino
sphere). The curves for various $r_\mathrm{wt}$ as a function of
$t_\mathrm{loc}$ obtained from equations~(4)-(6) are shown in Figure~1.

The wind trajectories from a hydrodynamical result by \citet[][$\sim
0.7-1.3$~s after bounce, gray lines]{Bura2006}, used in
\citet{Prue2006}, are compared with our models. Their wind
trajectories were obtained by mapping the two-dimensional model of an
exploding $15 M_\odot$ star to a one-dimensional grid at $\sim 0.5$~s
after bounce. In Figure~1, the time coordinate for each trajectory is
shifted to roughly match the one of our models. We find that their
model also exhibits a wind termination at $r\sim 500-1000$~km. The
temperature and density histories are, however, close to our models
with $r_\mathrm{wt} = 100-230$~km. This is due to their higher
neutrino luminosity ($\sim 2\times 10^{52}$~erg~s$^{-1}$) during the
relevant core-bounce time, a factor of two higher than assumed in our
models shown in Figure~1. This leads to a larger radius for a given
wind temperature (see \S~4.2 and Table~1).

The nucleosynthetic abundances in the neutrino-driven outflows are
calculated in a post-processing step by solving an extensive nuclear
reaction network code. The network consists of 6300 species between
the proton- and neutron-drip lines predicted by the recent fully
microscopic mass formula \citep[HFB-9,][]{Gori2005}, all the way from
single neutrons and protons up to the $Z = 110$ isotopes. All relevant
reactions, i.e. $(n, \gamma)$, $(p,\gamma)$, $(\alpha, \gamma)$, $(p,
n)$, $(\alpha, n)$, $(\alpha, p)$, and their inverses are
included. The experimental data, whenever available, and the
theoretical predictions for light nuclei ($Z < 10$) are taken from the
REACLIB\footnote{http://nucastro.org/reaclib.html.} compilation. All
the other reaction rates are taken from the Hauser-Feshbach rates of
BRUSLIB \footnote{http://www.astro.ulb.ac.be/Html/bruslib.html.}
\citep{Aika2005} making use of experimental masses \citep{Audi2003}
whenever available or the HFB-9 mass predictions \citep{Gori2005}
otherwise. The photodisintegration rates are deduced from the reverse
rates applying the reciprocity theorem with the nuclear masses
considered.

The $\beta$-decay rates are taken from the gross theory predictions
\citep[GT2;][]{Tach1990} obtained with the HFB-9 predictions
(T. Tachibana 2005, private communication). Electron capture reactions
on free nucleons and on heavy nuclei \citep{Full1982, Lang2001} as
well as rates for neutrino capture on free nucleons and $^4$He and for
neutrino spallation of free nucleons from $^4$He \citep{Woos1990,
  McLa1996} are also included. Neutrino-induced reactions of heavy
nuclei are not taken into account in this study, which are expected to
make only minor effects in this study.

Each nucleosynthesis calculation is initiated when the temperature
decreases to $T_9 = 9$, at which only free nucleons exist. The initial
compositions are then given by the electron fraction $Y_\mathrm{e, 9}$
(proton-to-baryon ratio) at $T_9 = 9$, such as $Y_\mathrm{e,9}$ and $1
- Y_\mathrm{e,9}$ for the mass fractions of free protons and neutrons,
respectively.

\begin{figure}
\epsscale{1.0}
\plotone{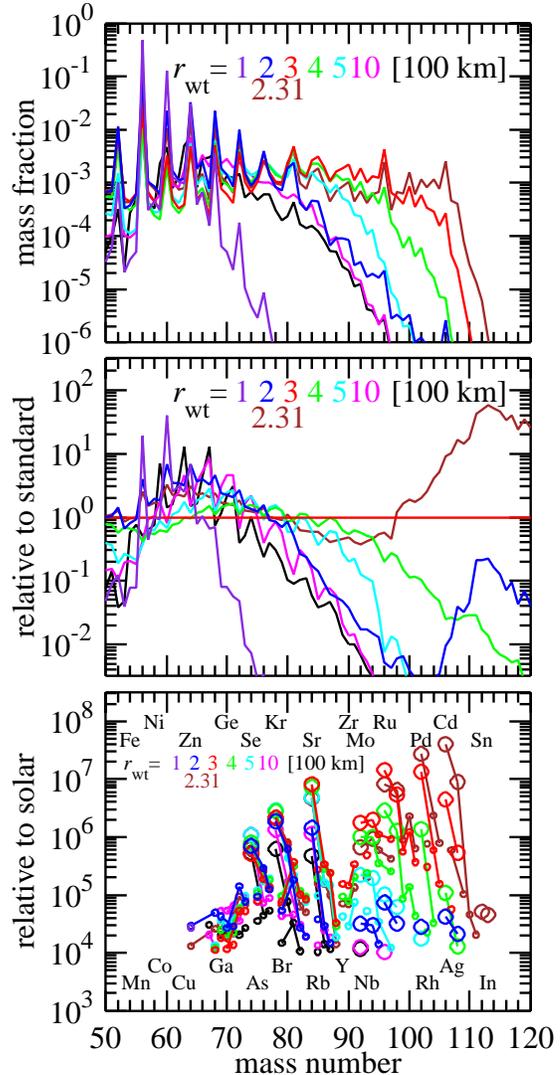}
\caption{Comparison of the nucleosynthetic results for various
 wind-termination radii $r_\mathrm{wt}$. The mass fractions
 (top) and their ratios relative to those for the standard model
 (middle) are shown 
 as a function of atomic mass number. The bottom  panel shows the abundances
 of isotopes (connected by a line for a given 
 element) relative to their solar values, where those lower than
 $10^4$ are omitted. The color coding corresponds to
 different values of $r_\mathrm{wt}$ as indicated in each panel (red
 is the standard model). The result for the outflow without
 wind termination is shown in black. In the 
 bottom panel, the names of elements are specified in the upper (even $Z$)
 and lower (odd $Z$) sides at their lightest mass numbers.}
\end{figure}

\begin{figure}
\epsscale{1.0}
\plotone{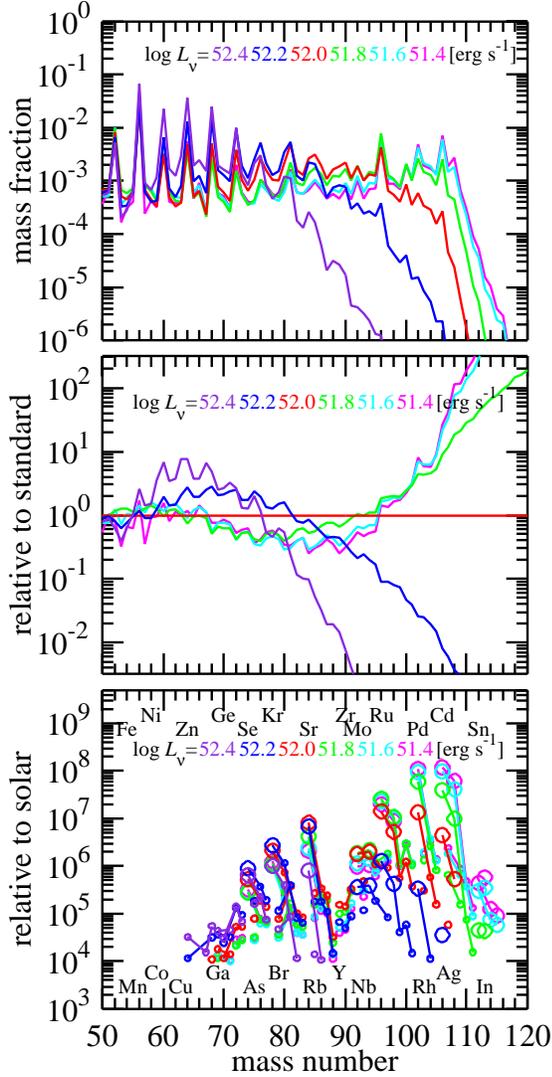}
\caption{Same as Figure~2, but for various neutrino luminosities ($L_\nu$).}
\end{figure}

\begin{figure}
\epsscale{1.0}
\plotone{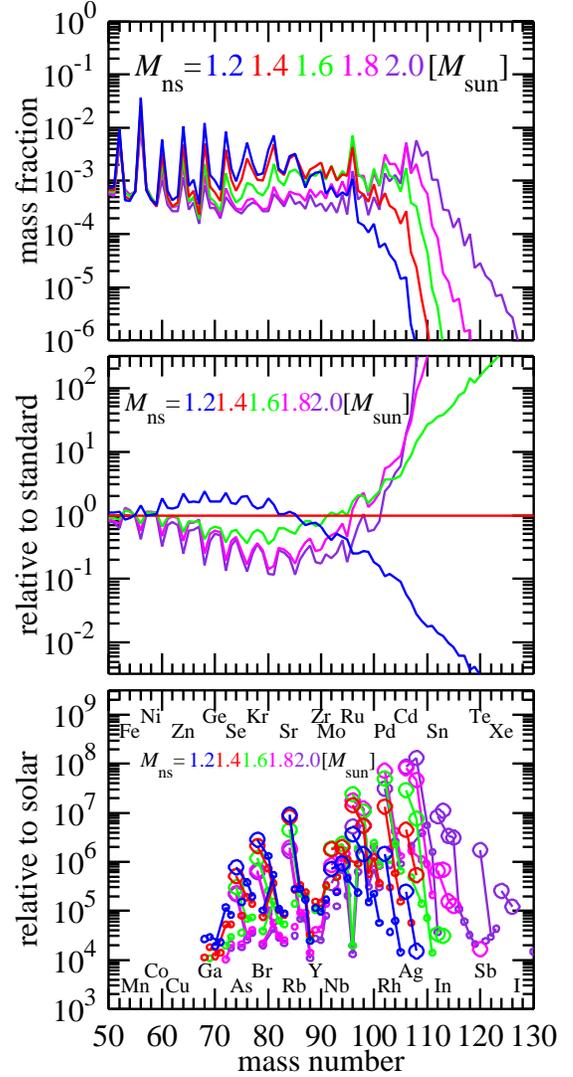}
\caption{Same as Figure~2, but for various neutron star masses ($M_\mathrm{ns}$).}
\end{figure}

\begin{figure}
\epsscale{1.0}
\plotone{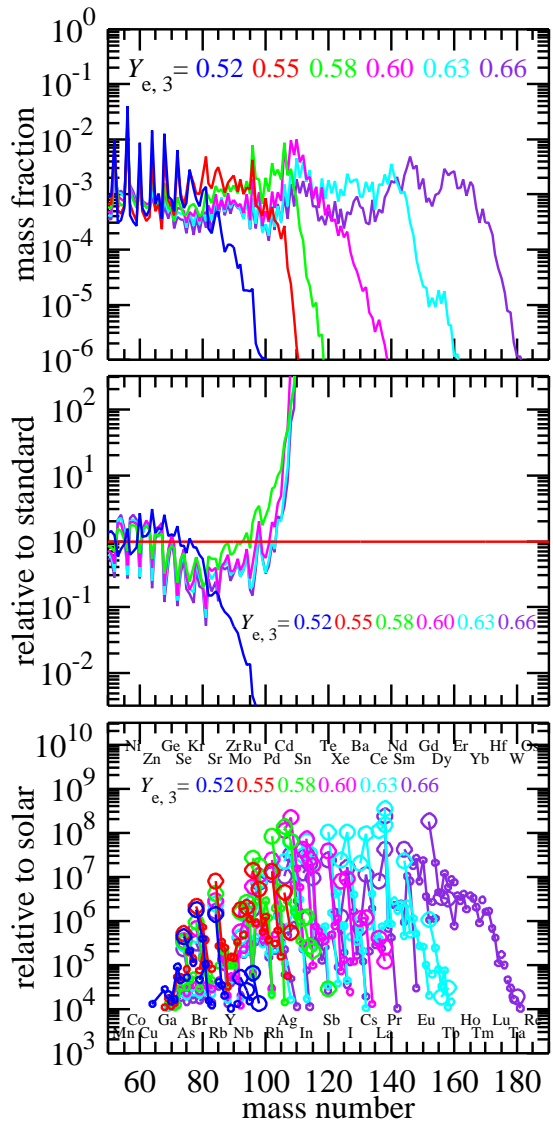}
\caption{Same as Figure~2, but for various electron fractions
 ($Y_\mathrm{e, 3}$).}
\end{figure}

\begin{figure}
\epsscale{1.0}
\plotone{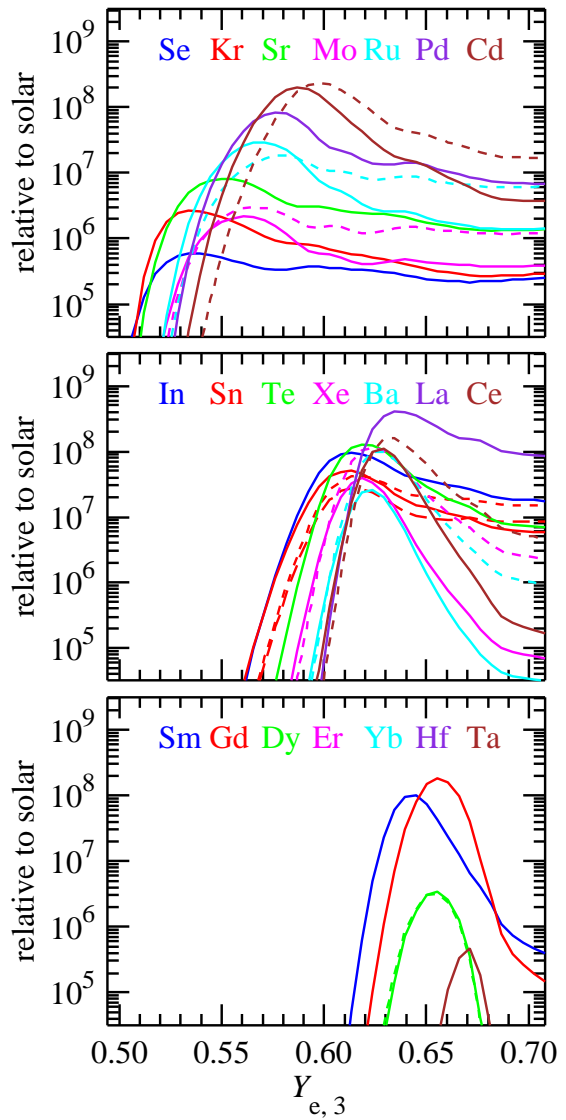}
\caption{Nucleosynthetic p-abundances relative to their solar values
 (production factors) as a function of $Y_\mathrm{e, 3}$.
 $M_\mathrm{ns}$, $L_\nu$, and
 $r_\mathrm{wt}$ are kept to be their fiducial values (1st line in
 Table~1). Each element is color coded with the solid, dashed, and
 long-dashed lines for the lightest,
 second-lightest, and third-lightest ($^{115}$Sn is only the case)
 isotopes, respectively (see 1st column of Table~4 for the list of p-nuclei).}
\end{figure}

\begin{figure}
\epsscale{1.0}
\plotone{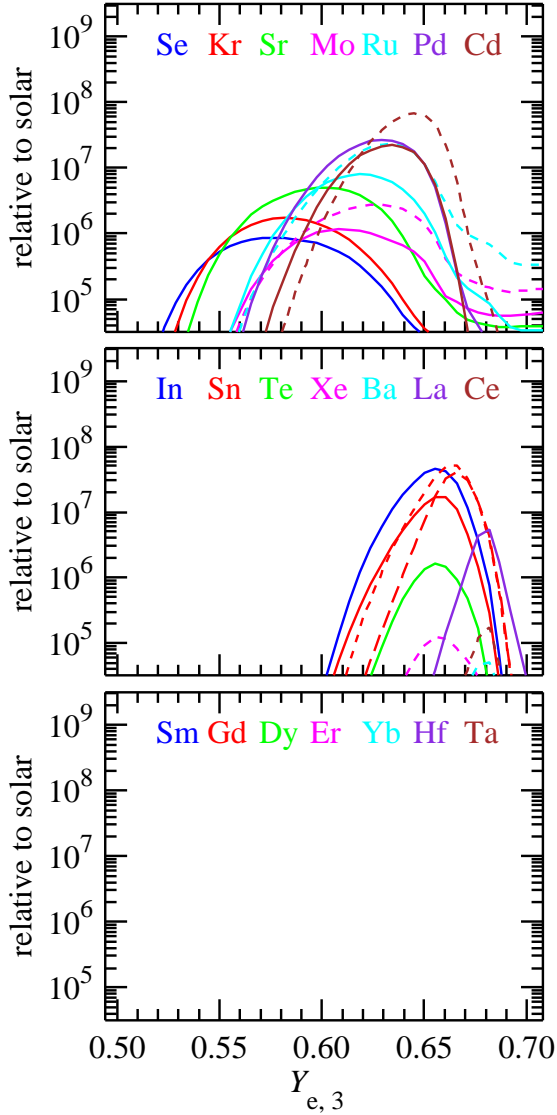}
\caption{Same as Figure~6, but for the case without wind termination
 ($r_\mathrm{wt} = \infty$).}
\end{figure}

\begin{figure*}
\epsscale{1.0}
\plotone{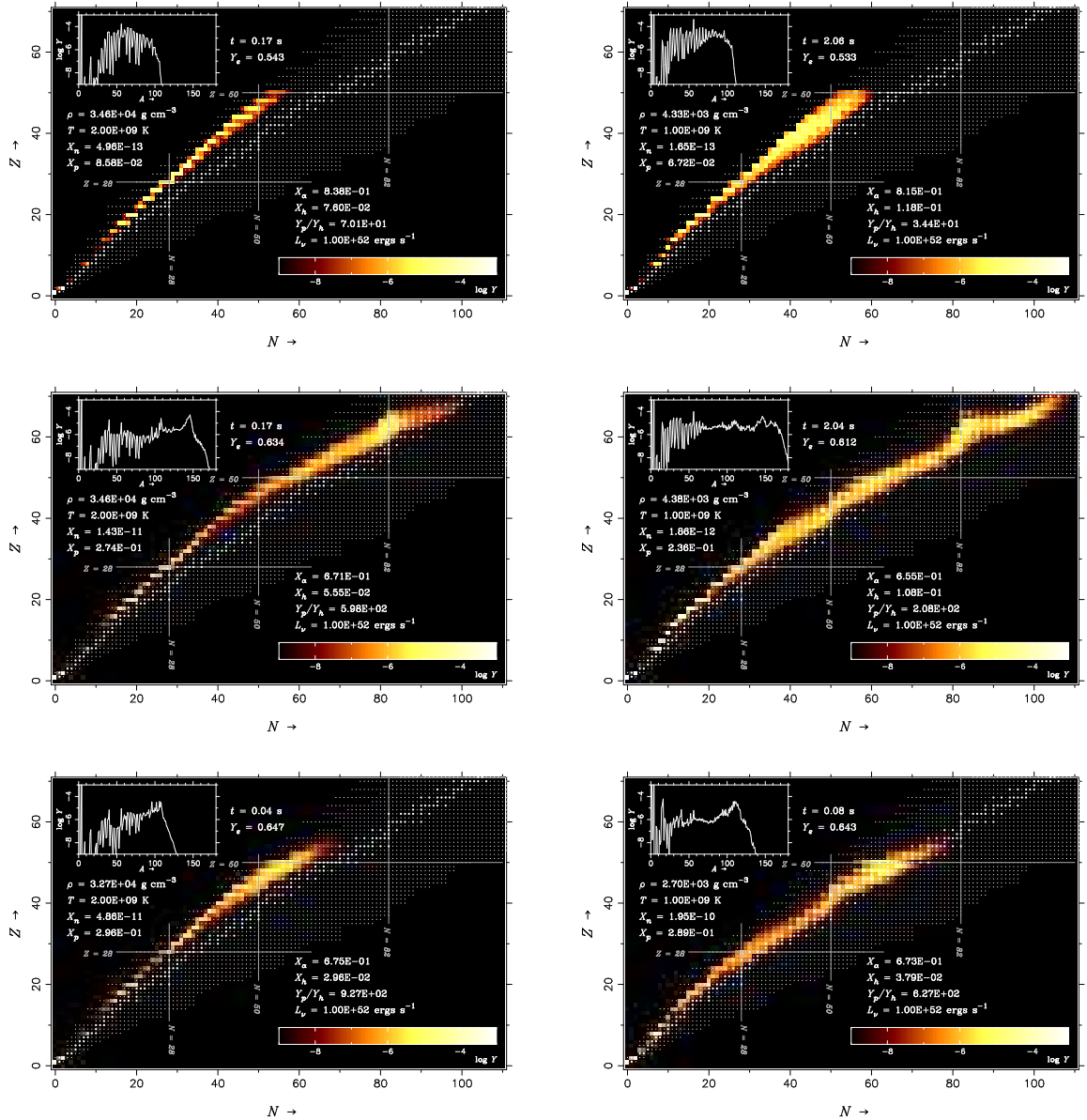}
\caption{Snapshots of nucleosynthesis on the nuclear chart when the
 temperature drops to $T_9 = 2$ (left) and 1 (right). Top, middle, and
 the bottom panels are for the standard model (1st line in Table~1),
 that with $Y_\mathrm{e, 9}$ replaced by 0.800 ($Y_\mathrm{e, 3} =
 0.655$), and that with $Y_\mathrm{e, 9}$ and $r_\mathrm{wt}$ replaced
 by 0.800 and $\infty$ (without wind termination). The nucleosynthetic
 abundances are color coded. The species included in the
 reaction network are shown by dots (with the thick dots for the stable
 isotopes). The abundance distribution as a function of atomic mass number is
 shown in the inset of each panel.}
\end{figure*}

\begin{figure}
\epsscale{1.0}
\plotone{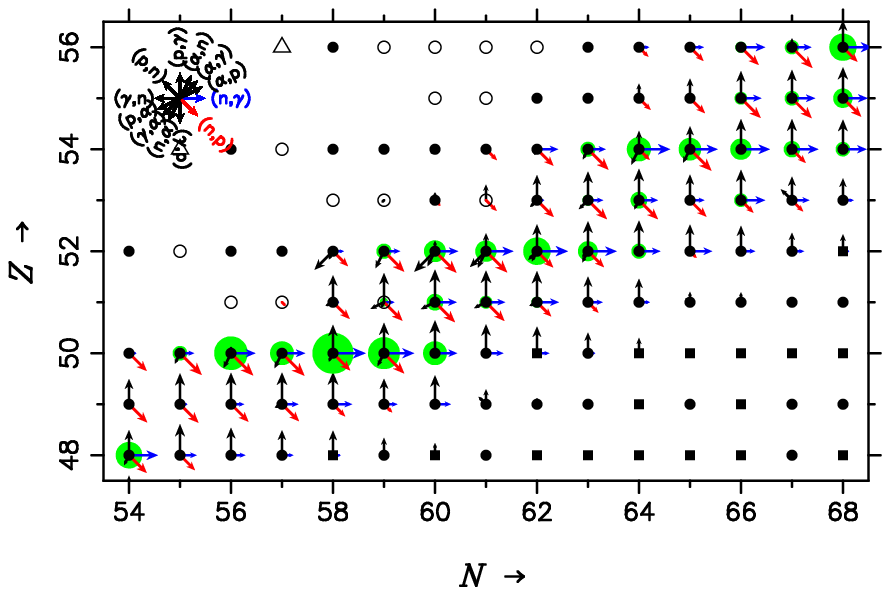}
\caption{Nuclear flows (arrows) and the abundances (circles) near the
  end point of the classical rp-process ($Z = 52$) in logarithmic
  scale for the model with $Y_\mathrm{e, 3} = 0.655$ (last line in
  Table~1) when the temperature decreases to $T_9 = 2$ (that
  corresponds to the middle-left panel in Figure~8).  The nuclei
  included in the reaction network are denoted by squares (stable
  isotopes), filled circles \citep[with measured masses
  of][]{Audi2003}, open circles \citep[with extrapolated masses
  of][]{Audi2003}, and triangles \citep[with the HFB-9 masses
  of][]{Gori2005}. The flows by $\beta^+$-decays (not shown here)
  are negligible compared to those by $(n, p)$ reactions (red
  arrows). Radiative neutron capture (blue arrows) also plays a
  significant role.  }
\end{figure}

\section{Uncertainties in Supernova Dynamics}

In the following subsections, we examine how the nucleosynthesis of
the $\nu$p-process is influenced by varying the wind-termination
radius $r_\mathrm{wt}$ (or temperature; \S~4.1), $L_\nu$ (\S~4.2),
$M_\mathrm{ns}$ (\S~4.3), and $Y_\mathrm{e, 9}$ (\S~4.4) from their
fiducial values 300~km (or $2.19\times 10^9$~K), $1\times 10^{52}\,
\mathrm{erg\ s}^{-1}$, $1.4\, M_\odot$, and 0.600, respectively, of
our standard model (1st line in Table~1). These values are taken as
those expected in the early wind phase of core-collapse
supernovae. All the explored models and their major outcomes are
summarized in Table~1 (the first 4 columns represent the input
parameters).

\subsection{Wind-termination Radius}

Recent hydrodynamic studies of core-collapse supernovae have shown that
the neutrino-driven outflows develop to be supersonic, which abruptly
decelerate by the reverse shock from the outer layers
\citep[e.g.,][]{Jank1995, Jank1996, Burr1995, Bura2006}. \citet{Arco2007} have
explored the effects of the reverse shock on the properties of
neutrino-driven winds by one-dimensional, long-term hydrodynamic
simulations of core-collapse supernovae. Their result shows that, in all
of their models ($10-25\, M_\odot$ progenitors), the outflows become
supersonic and form the termination shock when colliding with the slower
preceding supernova ejecta. This condition continues until the end of
their computations (10 seconds after core bounce) in their all of
``standard'' models with reasonable parameter choices. A recent
self-consistently exploding model of a $9\, M_\odot$ star also shows
qualitatively the same result \citep{Hued2010}.

In this subsection, we explore the effect of the wind-termination on the
$\nu$p-process. The termination point is located at $r_\mathrm{wt} =
100$, 200, 231, 300 (standard model), 400, 500, and 1000~km on the
transonic wind trajectory (black line) shown in Figure~1 (top
panel). The other parameters $L_\nu$, $M_\mathrm{ns}$, and $Y_\mathrm{e,
9}$ are kept to be the fiducial values (Table~1; 2nd to 9th lines).  In
Figure~1 (middle and bottom panels), we find shock-jumps of density and
temperature by wind termination only for the $r_\mathrm{wt} = 1000$~km
case, since the termination points are placed below the sonic radius
($r_\mathrm{s}= 515$~km; Figure~1, top panel) for the other
cases.\footnote{The outflows with $r_\mathrm{wt} < r_\mathrm{s}$ are
subsonic all the way. This happens in the early wind phase when the
slowly outgoing ejecta is still close to the core \citep{Arco2007}. In
this case, however, the mass ejection rate from the core is expected to
be close to that of the transonic solution (with the maximum $\dot
M$). Thus, the time variations of $r$, $\rho$, and $T$ may not be
substantially different from those of the transonic case for $r <
r_\mathrm{wt}$ \citep[see, e.g.,][]{Otsu2000}. We take, therefore, the
transonic solution for all the cases, rather than the subsonic solution
by introducing an additional free parameter $\dot M$.}

The result of nucleosynthesis calculations is shown in Figure~2. The top
panel shows the mass fractions, $X_A$, of nuclei as a function of atomic
mass number, $A$. We find that the case with $r_\mathrm{wt} = 231$~km
has the maximum efficiency of producing nuclei with $A = 100-110$
(including our calculations not shown here). The middle and bottom
panels show, respectively, the mass fractions relative to the standard
model ($= X_A/X_{A, \mathrm{standard}}$) and to their solar values
\citep{Lodd2003}, i.e., the production factor $f$ ($= X_i/X_{i,
\odot}$ for $i$-th isotope), as a function of $A$. We find a noticeable
effect of wind termination on the $\nu$p-process; the production of
p-nuclei between $A = 90$ and 110 is outstanding for the cases with
$r_\mathrm{wt} = 231$ and 300~km (standard model).

It should be noted that the asymptotic entropy $S$ ($= 57.0$ per
nucleon in units of the Boltzmann constant $k_\mathrm{B}$; Table~1) is
the same for all the cases here (except for $r_\mathrm{wt} = 1000$~km
owing to the termination-shock heating). These different outcomes can
be explained by the different values of $\Delta_\mathrm{n}$
($=0.24-17$, 13th column in Table~1), defined by equation~(1), owing
to the different expansion timescales after wind termination. As
indicated by the yellow band in Figure~1 (bottom panel), we find
substantial differences in the temperature histories before or during
the $\nu$p-process phase (defined as $T_9 = 1.5-3$).

We define two expansion timescales $\tau_1$ and $\tau_2$ (7th and 8th
columns in Table~1); the former is the time elapsed from $T_9 = 6$ to
$T_9 = 3$ and the latter from $T_9 = 3$ to $T_9 = 1.5$. These
represent the durations of the seed production and of the
$\nu$p-process, respectively. As can be seen in Figure~1 (bottom
panel), $\tau_1$ ($= 17.5$~ms) and thus the proton-to-seed ratio
$Y_\mathrm{p}/Y_\mathrm{h}$ ($= 124$) at $T_9 = 3$ are the same except
for the case with $r_\mathrm{wt} = 100$~km. Nevertheless, the
different values of $\tau_2$ and thus $\Delta_\mathrm{n}$ (see
equation~(1)) lead to the different efficiencies of the
$\nu$p-process. We find that $\Delta_\mathrm{n} \sim 10$ is needed for
an efficient production of p-nuclei with $A \sim 100$. This requires
the wind-termination at $T_\mathrm{wt, 9} \sim 2-3$ (in units of
$10^9$~K) to obtain $n_{\bar{\nu}_\mathrm{e}} \sim 0.1$
(equation~(2)). For the standard model ($r_\mathrm{wt} = 300$~km and
$T_\mathrm{wt, 9} = 2.19$), the maximum production factor
($f_\mathrm{max}$ in Table~1) is obtained at $^{96}$Ru
($\mathrm{nuc}(f_\mathrm{max})$ in Table~1), a daughter nucleus of
$^{96}$Pd ($N=50$) on the $\nu$p-process pathway. We have the optimal
production ($\log f_\mathrm{max} = 7.67$ at $^{106}$Cd) with
$T_\mathrm{wt, 9} = 2.65$ when the termination point is set to
$r_\mathrm{wt} = 231$~km.

In Table~1, the nuclide with the largest mass number $A_\mathrm{max}$
with $f > f_\mathrm{max}/10$ is also shown (e.g., $^{106}$Cd for the
standard model; $\mathrm{nuc}(A_\mathrm{max})$ in Table~1), which is
taken to be the largest $A$ of the p-nuclei synthesized by the $\nu
$p-process. Given that our standard model represents a typical
supernova condition, this implies that the $\nu$p-process can be the
source of the solar p-abundances up to $A \sim 110$ (see \S~6 for more
detail). However, this favorable condition is not robust against a
variation of $r_\mathrm{wt}$ (and thus $T_\mathrm{wt}$); the outflows
with $r_\mathrm{wt} = 200$~km ($T_\mathrm{wt, 9} =
2.95$)\footnote{Despite the largest $\Delta_\mathrm{n}$ ($=17.1$)
  among the various $r_\mathrm{wt}$ models, the $r_\mathrm{wt}
  =200$~km model ends up with inefficient $\nu$p-processing. This is
  due to $\Delta_\mathrm{n}$ defined for $T_9 \le 3$ (equation~(1)),
  while the maximal efficiency of $\nu$p-processing is obtained with
  $T_\mathrm{wt} = 2.65$ in this case.} and $r_\mathrm{wt} \ge 500$~km
($T_\mathrm{wt, 9} < 1.55$) end up with $A_\mathrm{max} = 84$
($^{84}$Sr; Table~1). Note that the outflow with $r_\mathrm{wt} =
1000$~km leads to a similar result as that without wind termination
(black line in Figure~2; $r_\mathrm{wt} = \infty$ in Table~1). This
indicates that the role of wind termination is unimportant for
$T_\mathrm{wt, 9} < 1.5$.

We find no substantial $\nu$p-processing for the outflow with
$r_\mathrm{wt} = 100$~km (Figure~2). This is due to the substantially
smaller $Y_\mathrm{e}$ at the beginning of the $\nu$p-process ($T_9 =
3$), $Y_\mathrm{e, 3} = 0.509$ (only slightly proton-rich), than those
for the other cases (0.550; Table~1). As a result,
$Y_\mathrm{p}/Y_\mathrm{h}$ at $T_9 = 3$ is only 1.78, resulting in a
small $\Delta_\mathrm{n}$ ($=0.24$) in spite of the largest
$n_{\bar{\nu}_\mathrm{e}}$ among the various $r_\mathrm{wt}$ models
presented here. It should be noted that $Y_\mathrm{e, 3}$ is always
lower than $Y_\mathrm{e, 9}$ ($= 0.600$ in the present cases). This is
due to a couple of neutrino effects. One is that the asymptotic
equilibrium value of $Y_\mathrm{e}$ in the non-degenerate matter
consisting of free nucleons, which is subject to neutrino capture, is
$Y_\mathrm{e, a} \approx 0.56$ \citep[see, e.g.,][]{Qian1996} with the
neutrino luminosities and energies taken in this study. Hence, the
value starts relaxing from $Y_\mathrm{e, 9}$ toward $Y_\mathrm{e, a}$
as soon as the calculation initiates. The other effect is due to the
continuous $\alpha$-particle formation ($T_9 < 7$) from
inter-converting free protons and free neutrons that is subject to
neutrino capture, which drives $Y_\mathrm{e}$ towards 0.5
\citep[``$\alpha$-effect'',][]{Meye1998}. In the $r_\mathrm{wt} =
100$~km case, the wind-termination takes place at high temperature
($T_\mathrm{wt, 9} = 5.19$) and thus the long $\tau_1$ ($= 359$~ms)
leads to the low $Y_\mathrm{e, 3}$ owing to the neutrino effects.

In summary, our exploration here elucidates a crucial role of wind
termination on the $\nu$p-process. On one hand, a fast expansion above
the temperature $T_9 \sim 3$ (more precisely, $T_9 = 2.65$ in the
considered conditions) is favored to obtain a high proton-to-seed
ratio at the onset of the $\nu$p-process. On the other hand, a slow
expansion below this temperature, owing to wind termination, is needed
to obtain $\Delta_\mathrm{n} \sim 10$ for efficient $\nu$p-processing.

We presume that the reason for somewhat different outcomes in previous
studies of the $\nu$p-process described in \S~1 is largely due to
their different behaviors of wind termination. The temperature
histories of trajectories taken by \citet[][an exploding $15 M_\odot$
star]{Prue2006}, similar to our models with $r_\mathrm{wt} = 100-230$
($T_\mathrm{wt, 9} = 2.7-5.2$), lead to the production of p-nuclei up
to $A \sim 100$. The reason of weak $\nu$p-processing in \citet[][a
$20 M_\odot$ explosion]{Froe2006} may be rather due to the moderate
proton-richness (up to $Y_\mathrm{e} \sim 0.54$) in their simulations
(see \S~4.4 and Figure~6). In contrast, negligible production of
p-nuclei in the electron-capture supernova of a $9 M_\odot$ star
\citep[][also S. Wanajo et al., in preparation]{Wana2009} is due to
the absence of a wind-termination shock within the relevant
temperature range ($T_9 = 1.5-3$) owing to the steep density gradient
of the oxygen-neon-magnesium core progenitors surrounded by a diluted
outer H/He envelope.

\subsection{Neutrino Luminosity}

The neutrino luminosity $L_\nu$ decreases with time from its initial
value of a few $10^{52}$~erg~s$^{-1}$ to $\sim 10^{51}$~erg~s$^{-1}$
during the first 10~s \citep{Fisc2010, Hued2010}. In this subsection, we
examine the effect of $L_\nu$ on the $\nu$p-process, by varying its
value from $10^{52.4} = 2.51 \times 10^{52}$~erg~s$^{-1}$ to 10 times
smaller than that with an interval of 0.2~dex (from 10th to 15th lines
in Table~1 and Figure~3). $M_\mathrm{ns}$ and $Y_\mathrm{e, 9}$ are
taken to be the fiducial values of $1.4\, M_\odot$ and 0.600,
respectively. In \S~4.1, we found that the temperature at the
wind-termination, $T_\mathrm{wt}$, plays a crucial role for the $\nu
$p-process. Hence, we adjust $r_\mathrm{wt}$ (Table~1) such that the
fiducial value of $T_\mathrm{wt, 9} = 2.19$ is obtained for each
$L_\nu$.

The results of nucleosynthesis calculations are shown in Figure~3 and
Table~1. We clearly see the increasing efficiency of $\nu$p-processing
with a decrease of $L_\nu$. This is due to the larger entropy for a
smaller $L_\nu$ (Table~1), while the expansion timescales $\tau_1$
(prior to the $\nu$p-process) are similar\footnote{When the radius of
  the neutrino sphere $R_\nu$ is fixed to a constant value, the
  expansion timescale increases with decreasing $L_\nu$\citep[see,
  e.g.][]{Otsu2000, Wana2001}. In this study, however, $R_\nu$ is
  assumed to decrease with decreasing $L_\nu$ (\S~2), which is more
  realistic. As a result, the difference of $\tau_1$ in the range of
  $L_\nu$ explored here is moderate.}. This leads to a higher
$Y_\mathrm{p}/Y_\mathrm{h}$ at the onset of the $\nu$p-process for a
lower $L_\nu$. In addition, the somewhat larger timescale $\tau_2$ for
a smaller $L_\nu$ increases $n_{\bar{\nu}_\mathrm{e}}$ (12th column in
Table~1). For these reasons, a smaller $L_\nu$ model achieves larger
$\Delta_\mathrm{n}$, leading to a more efficient $\nu$p-process.

It should be noted that in our explored cases, $r_\mathrm{wt}$ decreases
with decreasing $L_\nu$ (Table~1) in order to obtain the fiducial value
of $T_\mathrm{wt, 9} = 2.19$ (to figure out solely the effect of
$L_\nu$). However, if $r_\mathrm{wt}$ increases with time and thus
$T_\mathrm{wt}$ decreases with decreasing $L_\nu$, as in many explosion
models, only the early stage of the neutrino-driven wind with $L_\nu
\sim 10^{52}$~erg~s$^{-1}$ may be relevant to the high $T_\mathrm{wt, 9}
= 1.5-3$ \citep[see, e.g.,][]{Arco2007} that is needed for efficient $\nu
$p-processing (\S~4.1).

\subsection{Neutron Star Mass}

The mass of the proto-neutron star $M_\mathrm{ns}$ can be somewhat
different from its canonical value of $1.4\, M_\odot$, depending on its
progenitor mass. In this subsection, we examine the nucleosynthesis
calculations with $M_\mathrm{ns} = 1.2, 1.4, 1.6, 1.8$, and $2.0\,
M_\odot$, while $L_\nu$ and $Y_\mathrm{e, 9}$ are kept to be their
fiducial values of $10^{52}$~erg~s$^{-1}$ and 0.600. For each case, the
fiducial value of $T_\mathrm{wt, 9} = 2.19$ is obtained by adjusting
$r_\mathrm{wt}$ (from 16th to 20th lines in Table~1) as in \S~4.2.

We find a clear correlation between an increase of $M_\mathrm{ns}$ and
an increasing efficiency of $\nu$p-processing in Figure~4 and
Table~1. This is due to a larger $S$ and a smaller $\tau_1$ for a
larger $M_\mathrm{ns}$ \citep[e.g.,][]{Otsu2000, Wana2001}, both of
which help to increase $Y_\mathrm{p}/Y_\mathrm{h}$ and thus
$\Delta_\mathrm{n}$. This means that a more massive progenitor (up to
$\sim 30\, M_\odot$, which forms a neutron star) is favored for the
$\nu$p-process, given that all the other parameters are the same. In
reality, however, other factors, such as the evolutions of $L_\nu$,
$r_\mathrm{wt}$, and $Y_\mathrm{e}$ should be dependent on the
progenitor mass \citep[e.g.,][]{Arco2007}, which prevents us from
drawing any firm conclusions. It should be emphasized, however, that
the outflow with a typical mass of $M_\mathrm{ns} = 1.4\, M_\odot$ can
already provide physical conditions sufficient for producing the
p-nuclei up to $A \sim 110$.

\subsection{Electron Fraction}

The electron fraction $Y_\mathrm{e}$ is obviously one of the most
important ingredients in the $\nu$p-process as it controls the
proton-richness in the ejecta. Recent hydrodynamical studies with
elaborate neutrino transport indicate that $Y_\mathrm{e}$ exceeds 0.5
and increases up to $\sim 0.6$ during the neutrino-driven wind phase
\citep{Fisc2010, Hued2010}. It should be noted that $Y_\mathrm{e}$
substantially decreases from its initial value owing to the neutrino
effects (\S~4.1). In our standard model, the value decreases from
$Y_\mathrm{e, 9} = 0.600$ (at $T_9 = 9$) to $Y_\mathrm{e, 3} = 0.550$ at
the onset of the $\nu$p-process ($T_9 = 3$). However, these neutrino
effects would highly dependent on the neutrino luminosities and energies
of electron and anti-electron neutrinos assumed in this study. In this
subsection, therefore, we take the value at the onset of the $\nu
$p-process, $Y_\mathrm{e, 3}$, as a reference, rather than the initial
value $Y_\mathrm{e, 9}$.

Figure~5 and Table~1 (the last 6 lines) show the nucleosynthetic
results for $Y_\mathrm{e, 3} = 0.523, 0.550, 0.576, 0.603, 0.629$, and
0.655 (see Table~1 for their initial values $Y_\mathrm{e, 9}$). The
other parameters $M_\mathrm{ns}$, $L_\nu$, and $r_\mathrm{wt}$ (and
thus $T_\mathrm{wt}$) are kept to be their fiducial values (1st line
in Table~1). We find a great impact of the $Y_\mathrm{e}$ variation;
an increase of only $\Delta Y_\mathrm{e, 3} \sim 0.03$ leads to a
10-unit increase of $A_\mathrm{max}$, while $f_\mathrm{max}$ is
similar for $Y_\mathrm{e, 3} > 0.550$. This is due to the larger
$Y_\mathrm{p}/Y_\mathrm{h}$ (at $T_9 = 3$) for a larger $Y_\mathrm{e,
  3}$, leading the larger $\Delta_{n}$ despite the same
$n_{\bar{\nu}_\mathrm{e}}$ (Table~1).

In order to elucidate the effect of $Y_\mathrm{e}$ in more detail, the
production factor $f$ for each p-nucleus is drawn in Figure~6 as a
function of $Y_\mathrm{e, 3}$, where $M_\mathrm{ns}$, $L_\nu$, and
$r_\mathrm{wt}$ are kept to be their fiducial values. Each element is
color coded with the solid, dashed, and long-dashed lines for the
lightest, second-lightest, and third-lightest ($^{115}$Sn is only the
case) isotopes, respectively (see 1st column of Table~4 for the list of
p-nuclei). We find in the top panel of Figure~6 that the p-nuclei up
to $A = 108$ ($^{108}$Cd) take the maximum production factors
between $Y_\mathrm{e, 3} = 0.53$ and 0.60. Given the maximum
$Y_\mathrm{e, 3}$ to be $\sim 0.6$ according to some recent hydrodynamic
results \citep[e.g.,][]{Fisc2010, Hued2010}, this implies that the maximum
mass number of the p-nuclei produced by the $\nu$p-process is $A \sim
110$.

In principle, the heavier p-nuclei can be synthesized if the matter is
more proton-rich than $Y_\mathrm{e, 3} = 0.6$. The middle panel of
Figure~6 shows that the production factors of the p-nuclei from $A =
113$ ($^{113}$In) up to $A = 138$ ($^{138}$Ce) are maximal between
$Y_\mathrm{e, 3} = 0.61$ and 0.63. Furthermore, $^{144}$Sm and
$^{152}$Gd reach the maximum production factors at $Y_\mathrm{e, 3} =
0.64$ and 0.66, respectively (bottom panel in Figure~6). The end point
of the $\nu$p-process appears to be at $A \sim 180$ ($^{180}$Ta) in
our explored cases. It should be noted that the wind termination also
plays a crucial role as explored in \S~4.1. This is evident if we
compare Figures~6 and 7, where the latter is the result for
$r_\mathrm{wt} = \infty$. Without wind termination, more
proton-richness ($\Delta Y_\mathrm{e, 3} \sim 0.05$) is required for a
given p-nucleus to be produced, but with a substantially smaller
production factor. The p-nuclei heavier than $A = 140$ cannot be
produced at all without wind termination (bottom panel in Figure~7).

We can understand the reason for the above result from Figure~8, which
displays the snapshots of nucleosynthesis for selected cases on the
nuclear chart when the temperature drops to $T_9 = 2$ (left) and 1
(right). Top, middle, and bottom panels are for the standard model, that
with $Y_\mathrm{e, 9}$ replaced by 0.800 ($Y_\mathrm{e, 3} = 0.655$),
and that with $Y_\mathrm{e, 9}$ and $r_\mathrm{wt}$ replaced by 0.800
and $\infty$ (without wind termination), respectively. In the standard
model ($Y_\mathrm{e, 3} = 0.550$), the nuclear flow proceeds along the
proton-drip line and encounters the proton-magic number $Z = 50$ ($A
\sim 100-110$). There are $\alpha$-unbound nuclei of $^{106-108}$Te ($Z
= 52$) just above $Z = 50$ along the proton-drip line, which is the end
point of the classical rp-process \citep{Scha2001}. This is why the $\nu
$p-process stops at $A \sim 110$ for $Y_\mathrm{e, 3} \lesssim 0.6$.

As $Y_\mathrm{e, 3}$ exceeds 0.6, radiative neutron capture becomes
more important and competes with proton capture \citep{Prue2006,
  Wana2006}. This is due to the large amount of free protons
($Y_\mathrm{p}/Y_\mathrm{h} = 1130$ at $T_9 = 3$ for the middle panels
of Figure~8; the last line in Table~1) that release free neutrons
owing to neutrino capture ($\Delta_\mathrm{n} = 94.2$). As a result,
the nuclear flow detours the end point of the classical rp-process ($N
= 54-56$) at $Z = 52$ towards the larger atomic number through the
nuclei with $N > 60$, as can be seen in Figure~9. The stagnation of
the flow at the neutron-magic number $N = 82$ in the middle panels of
Figure~8 clearly shows the importance of neutron capture. The
concentration of nuclei at $N = 82$ leads to the large production
factors of the p-nuclei with $A = 130 - 150$ as seen in Figure~6. Note
that the p-nuclide $^{144}$Sm is located on the $N = 82$ line.

Beyond $N = 82$, the increasing atomic number and the decreasing
temperature inhibit further proton capture. Note that
$n_{\bar{\nu}_\mathrm{e}}' \sim 0.3 n_{\bar{\nu}_\mathrm{e}}$ in our
explored models (see equations~(1) and (2)). Thus, neutron capture
still continues at this stage. As a result, the nuclear flow
approaches the $\beta$-stability line and finally enters to the
neutron-rich region at $A \sim 160$ as seen in the middle-right panel
of Figure~8. Without wind termination (but with the same parameters
otherwise), however, the rapidly decreasing temperature does not allow
the nuclear flow to reach $N = 82$ as seen in the bottom panels of
Figure~8. This is the reason for the inefficiency of producing heavy
p-nuclei in Figure~7.

\begin{figure}
\epsscale{1.0}
\plotone{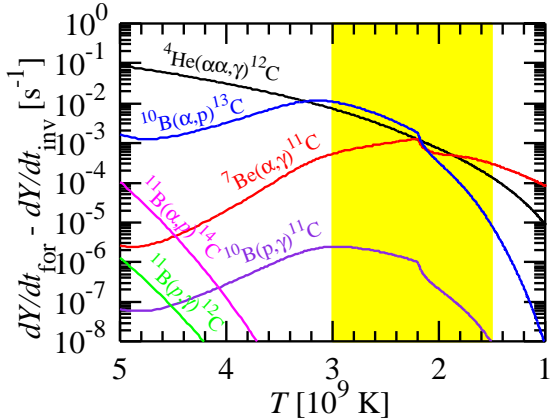}
\caption{Nuclear flows for the reactions that bridge from $A < 12$
  (the pp-chain region) to $A \ge 12$ as a function of
  temperature. The nuclear flow is defined as the deference of the
  time-derivatives (per second) of abundance between the forward and
  inverse reactions for a given channel. The yellow band indicates the
  temperature range relevant to the $\nu$p-process ($T_9 =
  1.5-3$).}
\end{figure}

\begin{figure}
\epsscale{1.0}
\plotone{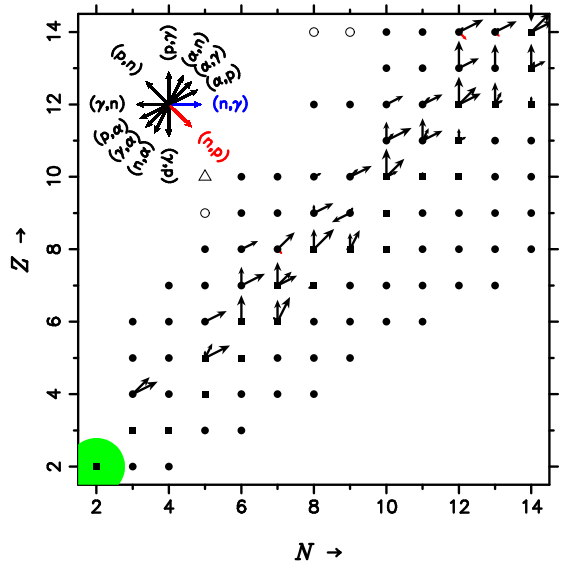}
\caption{Same as Figure~9, but for the standard model when the
  temperature decreases to $T_9 =2.5$, for a lighter $N$-$Z$
  region. At this temperature, the nuclear flows through
  $^7$Be$(\alpha, \gamma)^{11}$C$(\alpha, p)^{14}$N and
  $^7$Be$(\alpha, p)^{10}$B$(\alpha, p)^{13}$C play dominant roles for
  the breakout from the pp-chain region, along with the
  triple-$\alpha$ process (not shown here).}
\end{figure}

\begin{figure}
\epsscale{1.0}
\plotone{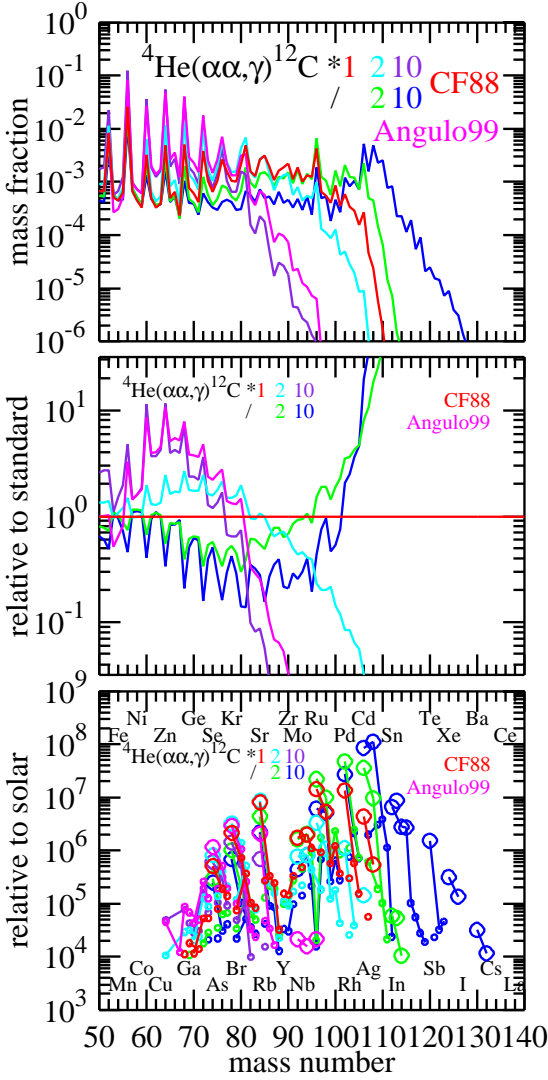}
\caption{Same as Figure~2, but for variations on the triple-$\alpha$ rate.}
\end{figure}

\begin{figure}
\epsscale{1.0}
\plotone{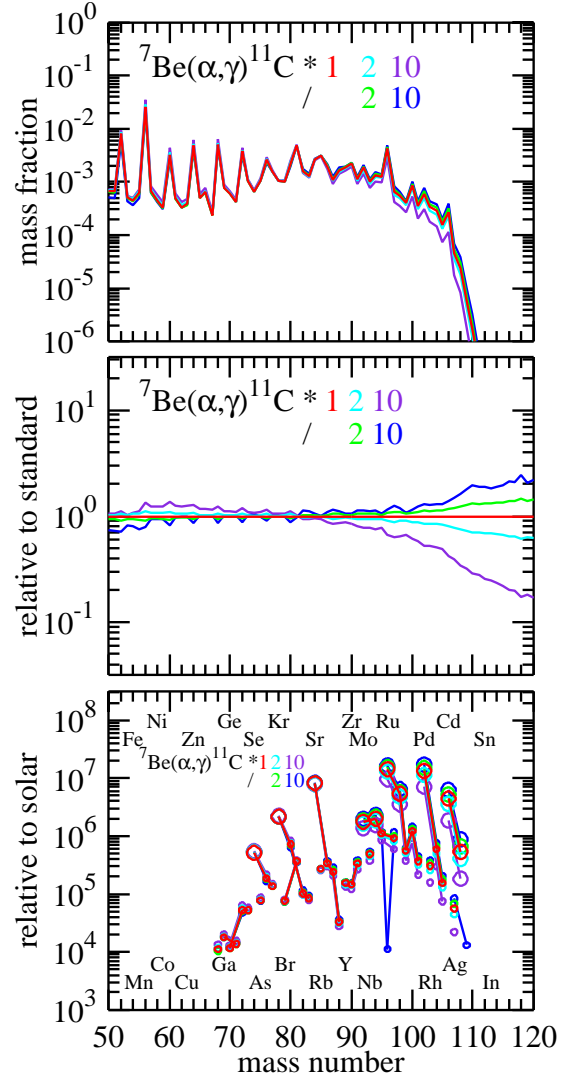}
\caption{Same as Figure~2, but for variations on the $^7$Be$(\alpha,
 \gamma)^{11}$C rate.}
\end{figure}

\begin{figure}
\epsscale{1.0}
\plotone{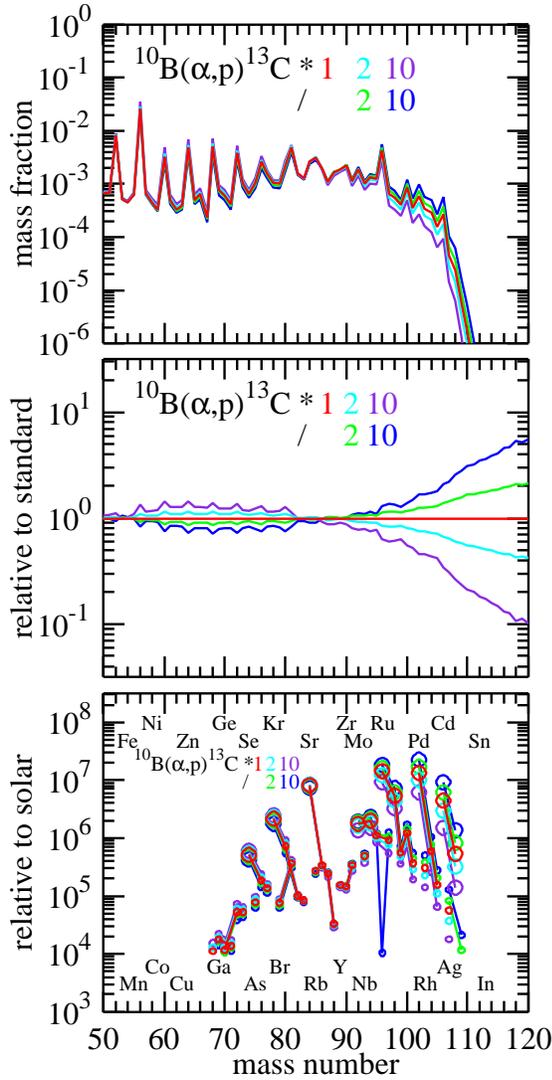}
\caption{Same as Figure~2, but for variations on the $^{10}$B$(\alpha,
 p)^{13}$C rate.}
\end{figure}

\begin{figure}
\epsscale{1.0}
\plotone{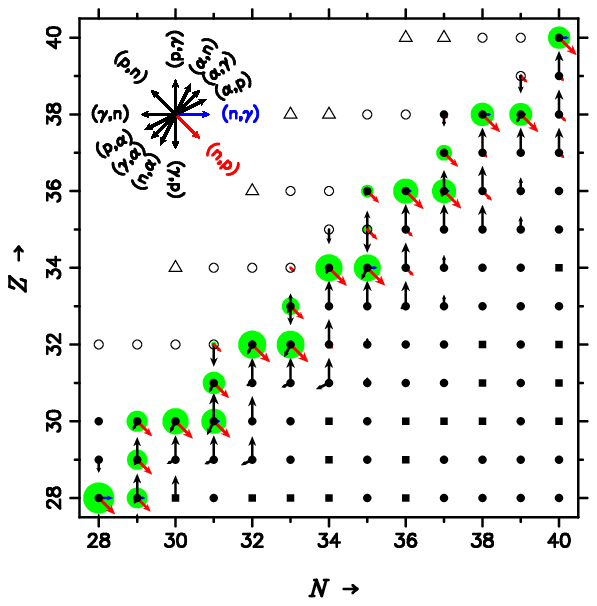}
\caption{Same as Figure~9, but for the standard model when the
  temperature decreases to $T_9 =2.0$.  }
\end{figure}

\begin{figure}
\epsscale{1.0}
\plotone{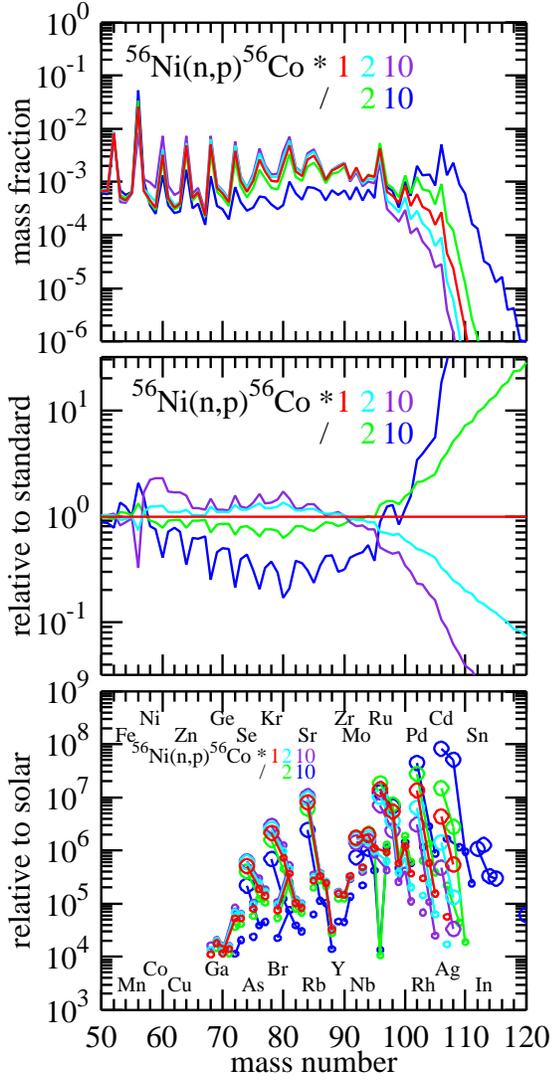}
\caption{Same as Figure~2, but for variations on the $^{56}$Ni$(n, p)^{56}$Co rate.}
\end{figure}

\begin{figure}
\epsscale{1.0}
\plotone{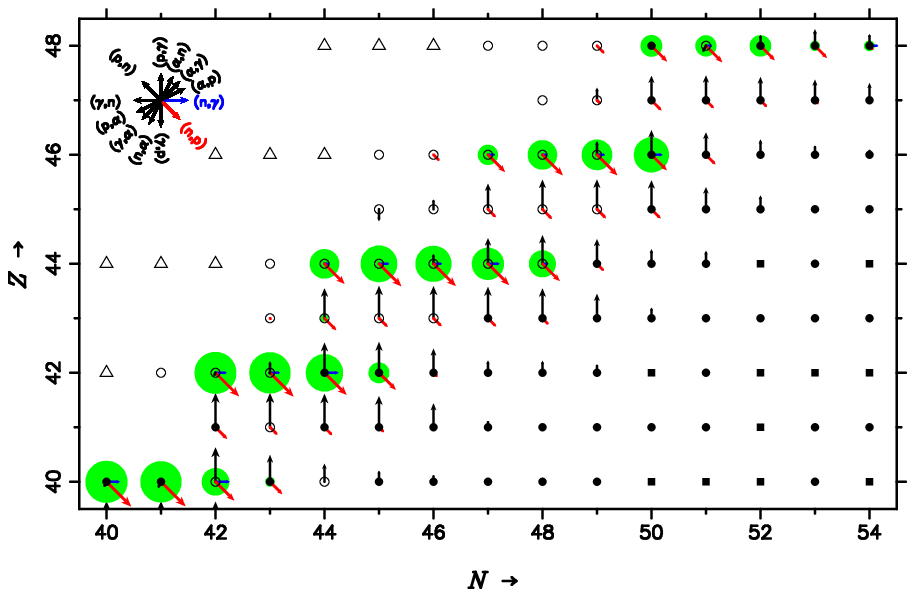}
\caption{Same as Figure~9, but for the standard model when the
  temperature decreases to $T_9 =2.0$.}
\end{figure}

\begin{figure}
\epsscale{1.0}
\plotone{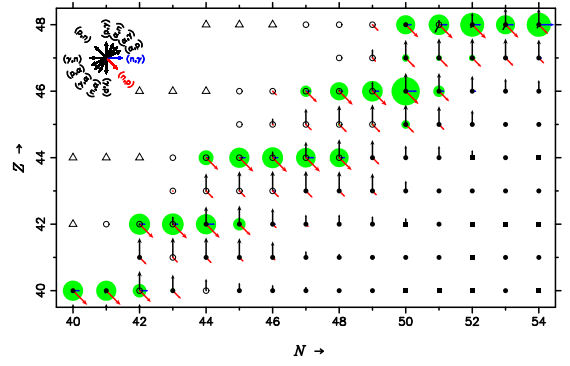}
\caption{Same as Figure~17, but for the model with the $^{56}$Ni$(n,
  p)^{56}$Co rate and its inverse reduced by a factor of 10 (at $T_9 =
  2.0$).}
\end{figure}

\begin{figure}
\epsscale{1.0}
\plotone{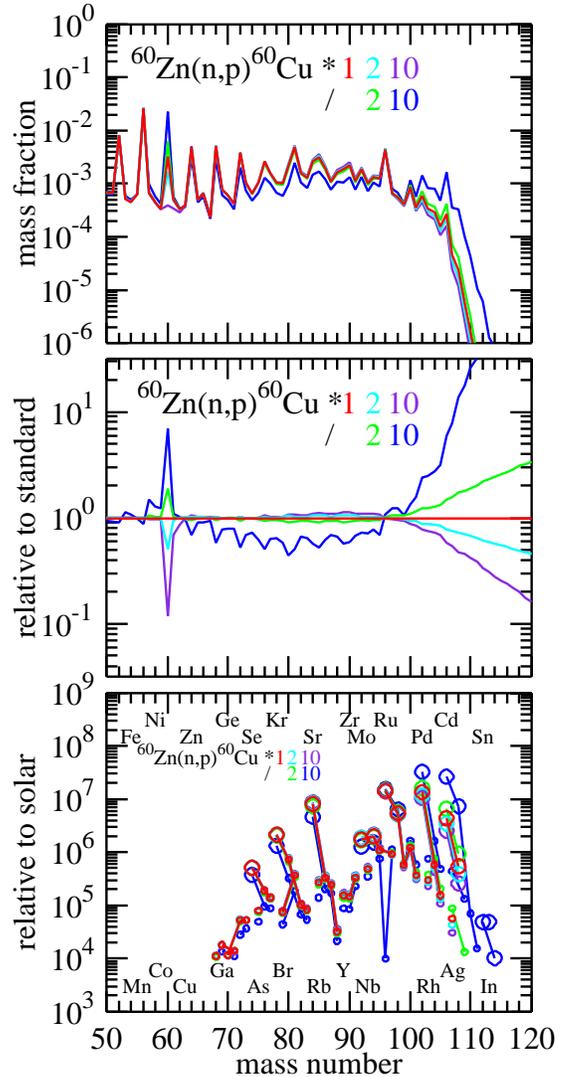}
\caption{Same as Figure~2, but for variations on the $^{60}$Zn$(n, p)^{60}$Cu rate.}
\end{figure}

\begin{figure}
\epsscale{1.0}
\plotone{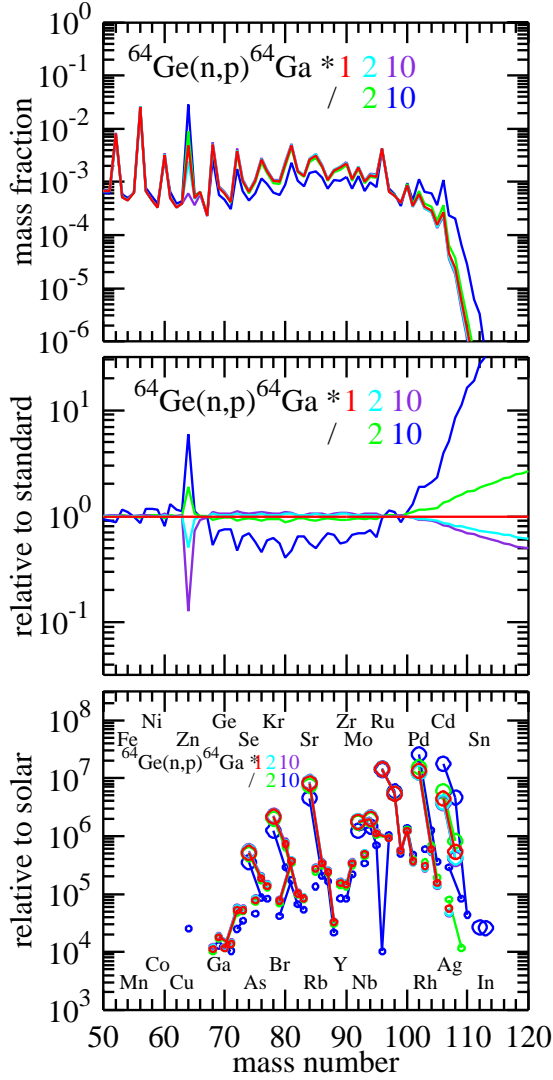}
\caption{Same as Figure~2, but for variations on the $^{64}$Ge$(n, p)^{64}$Ga rate.}
\end{figure}


\begin{figure}
\epsscale{1.0}
\plotone{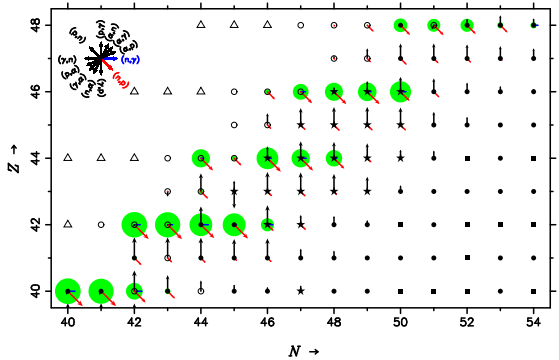}
\caption{Same as Figure~17, but for the standard model with the new
  experimental masses of \citet[][stars]{Webe2008}.  }
\end{figure}

\begin{figure}
\epsscale{1.0}
\plotone{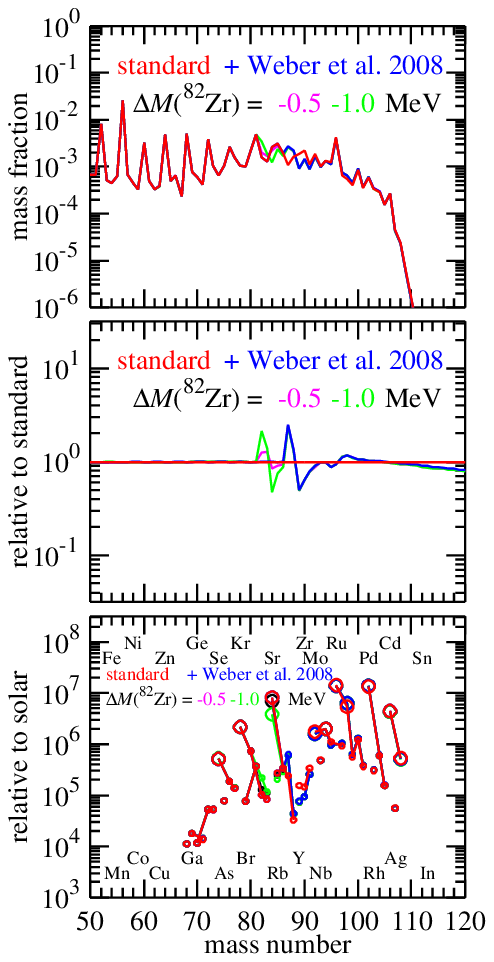}
\caption{Same as Figure~2, but for the standard model (red) with the
  new experimental masses of \citet[][blue]{Webe2008}. Also shown are
  the calculations with reductions ($-0.5$ and $-1.0$~MeV) of the
  $^{82}$Zr mass.}
\end{figure}

\section{Uncertainties in Nuclear Physics}

There have been continuing experimental works relevant to the $\nu
$p-process \citep[e.g.,][]{Webe2008, Haya2010} since its
discovery. However, we still rely upon theoretical or limited
experimental estimates for the vast majority of nuclear reactions
accompanied with the $\nu$p-process, which may suffer from
uncertainties. There are also a number of isotopes without
experimental mass measurements on the $\nu$p-process pathway
\citep{Webe2008}.

The $\nu$p-process is unique in the following aspects, different from
the classical rp-process. First is that the seed nuclei are directly
formed from free nucleons (i.e., the primary process), while the
classical rp-process needs CNO seeds. Thus, the triple-$\alpha$
process and some 2-body reactions relevant to the breakout from the
pp-chain region ($A < 12$) play important roles for setting the
proton-to-seed ratio $Y_\mathrm{p}/Y_\mathrm{h}$ (and thus
$\Delta_\mathrm{n}$) at the beginning of the $\nu$p-process
(\S~5.1). Second is the role of neutron capture, in particular of $(n,
p)$ reactions on heavy nuclei in the proton-rich matter, which bypass
the $\beta^+$-waiting points on the classical rp-process path
(\S~5.2). Third, the $\nu$p-process path is limited to $Z \le N$,
where most of the nuclear masses of relevance are measurable
\citep[][\S~5.3]{Webe2008}. This is an advantage compared to the
classical rp-process that proceeds through even-even $Z=N$ nuclei with
radiative proton capture to $Z > N$ isotopes \citep{Brow2002}.

In the following subsections \S~5.1 and 5.2, we test the effect of
uncertainties in some selected reactions by simply multiplying or
dividing their original values by factors of 2 and 10 with the
standard model (1st lines in Tables~1 and 3). All the explored results
are listed in Table~3. In \S~5.3, the effect of new mass measurements
by \citet{Webe2008} is discussed, along with possible uncertainties of
other unmeasured nuclear masses on the $\nu$p-process pathway.

\subsection{Breakout from the pp-Chain Region}

In Figure~10, the nuclear flows for the reactions that bridge from $A
< 12$ (the pp-chain region) to $A \ge 12$ (the CNO region) are shown
as a function of the temperature before ($T_9 > 3$) and after ($T_9 <
3$) the onset of the $\nu$p-process. The nuclear flows at $T_9 = 2.5$
for the relevant $N$-$Z$ region are also shown in Figure~11. Here, the
nuclear flow is defined as the difference between the time-derivatives
of abundances for the forward and inverse reactions of a given
channel. It is clear that, at a high temperature ($T_9 > 3$), the
triple-$\alpha$ process \citep[with the rate of][]{Caug1988} plays a
dominant role for the breakout from the pp-chain region. We find,
however, a couple of 2-body reaction sequences $^7$Be$(\alpha,
\gamma)^{11}$C$(\alpha, p)^{14}$N and $^7$Be$(\alpha,
p)^{10}$B$(\alpha, p)^{13}$C compete with the triple-$\alpha$ process
during the $\nu$p-process phase\footnote{$^7$Be$(\alpha, p)^{10}$B is
  an endothermic reaction. Because of its small (negative) $Q$-value
  of $-1.146$~MeV and the larger abundance of $\alpha$ particles, a
  small amount of $^{10}$B (that is immediately taken away by the
  subsequent $(\alpha, p)$ reaction) exists in the present case.}.

Table~2 lists the reaction rates and decay timescales for the relevant
isotopes at $T_9 = 2.5$ and 2.0. It is clear that $^7$Be$(\alpha,
\gamma)^{11}$C, 4 orders of magnitude slower than $^{11}$C$(\alpha,
p)^{14}$N, governs the former sequence. For the latter,
$^{10}$B$(\alpha, p)^{13}$C, although a factor of 10 smaller than
$^7$Be$(\alpha, p)^{10}$B, mainly controls the reaction flow, which
takes away nuclear abundances from $^{10}$B formed by the endothermic
reaction.

Figure~10 shows that $^7$Be$(\alpha, \gamma)^{11}$C and
$^{10}$B$(\alpha, p)^{13}$C exhibit similar roles to triple-$\alpha$
in the temperature range relevant to the $\nu$p-process. Therefore, we
select these three reactions for the sensitivity tests. Note that the
unstable isotope $^{11}$C produced is followed by $^{11}$C$(\alpha,
p)^{14}$N \citep[see][for a recent experimental evaluation of this
rate]{Haya2010} before decaying back to $^{11}$B.

All the data of these three reactions, from \citet[][for
$^{10}$B$(\alpha, p)^{13}$C]{Wago1969} and \citet[][for the
remainder]{Caug1988} in the REACLIB compilation, are based on
experimental information of single resonance states. Contribution from
(possible) resonances at higher excitation energies could thus sizably
change these rates. As an example, the triple-$\alpha$ rate of
\citet{Angu1999}, which includes contribution from the 9.2-MeV $2^+$
state that is predicted theoretically, leads to a factor of 2 to 10
higher values (for the temperature range relevant to seed production,
$T_9 = 7-3$) than that of \citet{Caug1988} based on the single 7.6-MeV
$0^+$ (Hoyle) state. Recent experimental works did not confirm the
presence of the 9.2~MeV state, but other levels in this energy region
as well as those at higher energies might contribute to this rate
\citep{Aust2005, Dige2005, Dige2009}.

The result of sensitivity tests for the triple-$\alpha$ rate is shown
in Figure~12, where the forward and inverse rates are multiplied or
divided by factors of 2 and 10. We find substantial changes in the
production of p-nuclei with $A \sim 100-110$ for a factor of 2
variation on the rate, and more drastic changes for a factor of 10
variation. It can be mainly attributed to the resulting proton-to-seed
ratio $Y_\mathrm{p}/Y_\mathrm{h}$ (at $T_9 = 3$) and thus
$\Delta_\mathrm{n}$ (3rd and 4th lines in Table~3). Note that
$n_{\bar{\nu}_\mathrm{e}}$ ($= 0.0834$; 1st line in Table~1) remains
the same for all the cases here. A larger triple-$\alpha$ rate leads
to a more efficient seed production and thus smaller
$Y_\mathrm{p}/Y_\mathrm{h}$ and $\Delta_\mathrm{n}$. A larger rate
during the $\nu $p-process phase ($T_9 = 1.5-3$) also yields more
carbon and other intermediate-mass nuclei that act as \textit{proton
  poison}. As a result, efficiency of the $\nu$p-process for heavy
element synthesis decreases. The same interpretation is applicable to
the opposite case with a smaller rate.

We find that a replacement of the triple-$\alpha$ rate by that of
\citet{Angu1999} inhibits production of p-nuclei for $A > 80$
(Figure~12). In fact, the net effect of including the 9.2~MeV state
(not confirmed by recent experiments) by \citet{Angu1999} corresponds
to the rate of \citet{Caug1988} multiplied by a factor of 10. This
demonstrates the importance of future re-evaluations of (possible)
contribution from higher levels than the 7.64~MeV state in $^{12}$C.

Figures~13 and 14 show the result for $^7$Be$(\alpha, \gamma)^{11}$C
and $^{10}$B$(\alpha, p)^{13}$C. We find non-negligible differences in
the p-abundances with $A \sim 100-110$, although the impact is much
smaller than that for triple-$\alpha$. Note that a larger rate has a
stronger impact than a smaller rate (middle panels). This is a
consequence of the fact that the larger rate of a given channel
increases the total efficiency for the breakout from the pp-chain
region, while the other two channels are still active for the smaller
rate (see Figure~10).

$^7$Be$(\alpha, \gamma)^{11}$C competes with triple-$\alpha$ only
during the late phase of the $\nu$p-process ($T_9 \lesssim 2$;
Figure~10). A larger rate during this phase leads to more production
of intermediate-mass nuclei that act as proton poison. A variation of
this rate does not substantially affect $Y_\mathrm{p}/Y_\mathrm{h}$
(at $T_9 = 3$) and $\Delta_\mathrm{n}$ at the onset of
$\nu$p-processing (Table~3). $^{10}$B$(\alpha, p)^{13}$C however
competes with triple-$\alpha$ at $T_9 \sim 2-3.5$ (Figure~10). Hence,
a variation on the rate also affects $Y_\mathrm{p}/Y_\mathrm{h}$ at
the beginning of $\nu$p-processing and $\Delta_\mathrm{n}$ (Table~3).

\subsection{$(n,p)$ Reactions on Heavy Nuclei}

The $\nu$p-process starts at $T_9 \sim 3$ from the seed nucleus
$^{56}$Ni, which is formed from free nucleons earlier. During the
$\nu$p-process, the $(n, p)$ reactions play an important role for
determining the nuclear flows. Figure~15 shows the nuclear flows
starting from $^{56}$Ni ($Z=N=28$) up to $^{80}$Zr ($Z=N=40$). The
isotopes included in the reaction network are denoted by squares
(stable isotopes), filled circles \citep[with measured masses
of][]{Audi2003}, open circles \citep[with extrapolated masses
of][]{Audi2003}, and triangles \citep[with the HFB-9 masses
of][]{Gori2005}. We find that the nuclear flow of the $\nu$p-process
proceeds through even-even $Z=N$ isotopes up to $Z=N=40$ as in the
classical rp-process. All the nuclear masses on the $\nu$p-process
path (up to $Z=N=40$), which determine the abundance distribution for
given isotones, were measured by experiments
\citep{Audi2003}\footnote{We do not take the mirror-mass evaluations
  of \citet{Brow2002} for $Z > N$ nuclei into account, as the
  $\nu$p-process path is limited to the $Z \le N$ region (except for a
  flow to $^{59}$Zn but with measured masses; Figure~15).}.

Currently, there are no experiment-based estimates for the $(n, p)$
reactions on proton-rich isotopes along the $\nu$p-process path. We
rely upon the theoretically predicted Hauser-Feshbach rates, which are
generally considered to involve uncertainties up to a factor of a few
\citep[this reduces to $\sim 40\%$ if the nuclear levels are well
determined and the level densities are large
enough,][]{Raus1997}. \citet{Raus2010} also finds sizable shifts of
effective energy windows for $(n, p)$ at high temperature, which might
modify these rates. In this subsection, therefore, the sensitivity
tests for $(n, p)$ reactions are made with factors of 2 and 10
variations as in \S~5.1.

We here pick up three $(n, p)$ reactions starting from the seed nuclei
along the $\nu$p-process path, i.e., on $^{56}$Ni, $^{60}$Zn, and
$^{64}$Ge. The last one, $^{64}$Ge, is the first $\beta^+$-waiting
point nucleus encountered in the classical rp-process path. Note that
the variations on these rates do not affect
$Y_\mathrm{p}/Y_\mathrm{h}$ at the onset of the $\nu$p-process nor
$\Delta_\mathrm{n}$ (Table~3). All these rates are from theoretical
estimates in BRUSLIB \citep{Aika2005} making use of experimental
masses \citep{Audi2003}. Our test calculations with the $(n, p)$ rates
replaced by those in the REACLIB compilation \citep{Raus2002} are in
reasonable agreement (within factor of a few) with our standard case
\citep[see also][]{Wana2009}.

We find a remarkable change in the p-abundances with $A \sim 110$ by a
factor of 10 with only a factor of 2 variation on $^{56}$Ni$(n,
p)^{56}$Co (Figure~16 and Table~3). This demonstrates that the $(n,
p)$ reaction on the first $(n, p)$-\textit{waiting nucleus} $^{56}$Ni
plays a key role for the progress of nuclear flows.

It should be noted that a smaller rate leads to more efficient
$\nu$p-processing as can be seen in the bottom panel of Figure~16 and
in Table~3 (see $f_\mathrm{max}$ and $A_\mathrm{max}$). The reason can
be explained as follows: Figure~17 extends the nuclear flows in
Figure~15. We find that the $\nu$p-process path proceeds through
even-even $Z=N$ isotopes and deviates from $^{84}$Mo ($Z=N=42$) toward
$Z < N$, reaching $^{96}$Pd on the $N=50$ shell closure. These
isotopes have large abundances during the whole $\nu$p-process phase,
as can be seen in Figures~15 and 17 (filled green circles). The top
panel of Figure~16 shows that the abundances with $A = 60, 64, 68, 72,
76, 80$, and 84 are similar to that of $A = 96$. Table~2 lists the
$(n, p)$ rates and decay timescales for the corresponding isotopes at
$T_9 = 2.5$ and 2.0. The $(n, p)$ rates on $^{56}$Ni and $^{96}$Pd,
neutron magic nuclei on $N = 28$ and 50, respectively, are a factor of
4--10 times smaller than the others. This indicates that the free
neutrons created by neutrino capture ($\Delta_\mathrm{n}$) are
preferentially consumed by the even-$Z$ nuclei with $30 \le Z \le 42$
(that act as \textit{neutron poisons}), rather than by $^{96}$Pd.

A reduction of the $^{56}$Ni$(n, p)$ rate (by a factor of 2 or 10)
reduces the abundances of these neutron poisons by a similar factor
($A \sim 60-90$, Figure~16; middle). Despite this, the abundance of
$^{96}$Pd does not decrease (even increases). This is due to the
faster $(n, p)$ rates for the $Z=N$ ($=30-42$) nuclei, causing the
nuclear flows from $^{56}$Ni to immediately reach $^{96}$Pd and to
stagnate there. As a result, a larger number of free neutrons becomes
available for the $^{96}$Pd$(n, p)$ reaction.

The nuclear flows for the $^{56}$Ni$(n, p)$ rate reduced by a factor
of 10 are shown in Figure~18. Smaller abundances of $Z=N$ nuclei
$^{80}$Zr and $^{84}$Mo can be seen, which leads to the larger flows
beyond $N=50$ through $^{96}$Pd. This clearly demonstrates that
$^{96}$Pd plays a role as a ``second seed nucleus'' for producing
nuclei heavier than $A = 96$. In short, a reduction of the
$^{56}$Ni$(n, p)$ rate increases the number of free neutrons available
for the second seed nuclei of $^{96}$Pd.

Figures~19 and 20 show the results for the second and third $(n,
p)$-waiting nuclei, $^{60}$Zn and $^{64}$Ge (the first
$\beta^+$-waiting nucleus on the classical rp-process). The variations
on these rates also lead to visible changes in the nucleosynthetic
p-abundances, being however less prominent than in the case of
$^{56}$Ni. Note that a reduced $(n, p)$ rate leads to a larger impact
on the nucleosynthetic p-abundances than an increased value of this
rate. This is due to the fact that the $(n, p)$ reaction on $^{56}$Ni
is substantially slower than those on $^{60}$Zn and $^{64}$Ge
(Table~2), where the strength of the nuclear flow is limited by the
former reaction.

\subsection{Nuclear masses on the $\nu$\lowercase{p}-process pathway}

Nuclear masses on the nucleosynthetic path are fundamental for all the
relevant nuclear (or weak) processes. In particular, the flow strength
of radiative proton capture during $\nu$p-processing ($T_9 = 3-1.5$)
is mostly determined from proton separation energies, where $(p,
\gamma) \leftrightarrow (\gamma, p)$ is generally faster than $(n, p)$
and $(n, \gamma)$ and thus in a quasi equilibrium. This explains the
concentration of abundances on even-$Z$ isotopes in Figures~8 (left
panel; in particular for $Z \le 50$), 9, 15, 17, and 18.

There are a number of isotopes without measured masses in the
compilation of \citet{Audi2003} for $40 \le Z \le 50$ (denoted by open
circles in Figures~17 and 18), including the parent nuclei of light
p-nuclei, $^{84}$Sr, $^{92, 94}$Mo, and $^{96,
  98}$Ru. \citet{Prue2006} noted that the unmeasured masses of
$^{92}$Ru and $^{93}$Rh (i.e., the proton separation energy of
$^{93}$Rh) on the $N=48$ isotones are crucial for determining the
ratio of $^{92}$Mo/$^{94}$Mo.

Recently, \citet{Webe2008} obtained precision measurements of a number
of nuclear masses along the $\nu$p-process pathway, including those of
$^{92}$Ru and $^{93}$Rh. Here, we present the nucleosynthetic result
with inclusion of there new masses, denoted by star symbols in
Figure~21, with our standard model (first lines in Table~1 and 3). We
confirm the suppression of the flow through $^{87}$Mo$(p,
\gamma)^{88}$Tc ($N=45$; see Figure~17), which has been reported in
\citet{Webe2008}. This leads to a factor of three enhancement of
$^{87}$Sr and a factor of two reduction of $^{89}$Y, which are however
not p-isotopes (and with small production factors). The other
p-abundances, including of $^{92, 94}$Mo, are almost unchanged, as
reported in \citet{Webe2008} and in \citet{Fisk2009}.

Our calculations of sensitivity tests for all other unmeasured masses
on the $\nu$p-process path show that the mass of $^{82}$Zr (or the
proton separation energy of $^{83}$Nb) on $N=42$ plays an important
role for production of a light p-nuclei $^{84}$Sr. The others have
only minor roles for the sensitivity tests with variations of up to
$\pm 1$~MeV on the nuclear masses. We find in Figure~22 that a 1.0~MeV
reduction of the $^{82}$Zr mass (equivalent to a reduction of the
proton separation energy of $^{83}$Nb) leads to a reduction of the
$^{84}$Sr abundance by a factor of two (middle panel). An increase of
the $^{82}$Zr mass has no effect on the p-abundances.

This is particularly important when we consider the role of the
$\nu$p-process to the solar inventory of most mysterious p-nuclei,
$^{92}$Mo and $^{94}$Mo. As can be seen in Figure~22 (bottom panel;
and in other similar figures), the production factors of $^{92}$Mo and
$^{94}$Mo are always substantially smaller than the neighboring
p-isotopes, in particular, $^{84}$Sr (see Figures~6 and 7). A
reduction of the $^{84}$Sr would in part reduce this large gap. Note
that the experimental mass of $^{93}$Nb in \citet{Audi2003} involves a
large uncertainty (315~keV). Future precision measurements of both
$^{82}$Zr and $^{83}$Nb are thus highly desired.

\section{$\nu$\lowercase{p}-process as the origin of \lowercase{p}-nuclei}

In the previous sections (\S~4 and 5), we find that uncertainties in
both the supernova dynamics and nuclear reactions can substantially
affect the productivity of p-nuclei. This makes it difficult to
determine the role of the $\nu$p-process as the source of the solar
p-nuclei. Keeping such uncertainties in mind, we discuss a possible
contribution of the $\nu$p-process to the solar p-abundances based on
our result by comparing with other possible sources.

Table~4 lists the currently proposed astrophysical origins for each
p-nuclide (1st column) with its solar abundance and fraction relative
to its elemental abundance \citep[2nd and 3rd
columns,][]{Lodd2003}. All these sources are associated with
core-collapse supernovae. Photo-dissociation of pre-existing
neutron-rich abundances in the oxygen-neon layer of core-collapse
supernovae (or in their pre-collapse phases), i.e., the
$\gamma$-process \citep{Woos1978, Pran1990, Raye1995, Raus2002,
  Haya2008} is currently regarded as the most successful scenario. In
the 4th column of Table~4, the p-nuclei whose origins can be explained
by the $\gamma$-process in \citet{Raye1995} are specified by
``yes''. The bracketed ones are those underproduced in a more recent
work by \citet{Raus2002}. The origins of up to 24 out of 35 p-isotopes
can be explained by the $\gamma$-process. However, the light
p-isotopes ($^{92, 94}$Mo, $^{96, 98}$Ru, $^{102}$Pd, $^{106, 108}$Cd,
$^{113}$In, and $^{115}$Sn), which account for a large fraction in the
solar p-abundances, and some heavy p-isotopes ($^{138}$La and
$^{152}$Gd) need other sources (specified by ``no'' in Table~4).

The $\nu$-process \citep[5th column in Table~4,][]{Woos1990} in
core-collapse supernovae is suggested to account for the production of
a couple of heavy p-isotopes $^{138}$La and $^{180}$Ta (the former is
underproduced in the $\gamma$-process). The $\alpha$-rich and slightly
neutron-rich ($Y_\mathrm{e} \approx 0.47-0.49$; slightly more
proton-rich than the $\beta$-stability values) neutrino-driven
outflows were also suggested as the production site of some light
p-isotopes including $^{92}$Mo \citep[but not $^{94}$Mo,][]{Hoff1996,
  Wana2006, Wana2009}. The proton-richness relative to the
$\beta$-stability line in the fragmented QSE clusters \citep{Hoff1996,
  Meye1998b} at $T_9 \sim 4-3$ leads to the formation of these
p-nuclei with $N \le 50$. Such QSE clusters on the proton-rich side of
the $\beta$-stability line will be denoted as ``p-QSE'' hereafter. A
recent study of nucleosynthesis in the electron-capture supernovae of
a $9\, M_\odot$ star shows that the lightest p-nuclei $^{74}$Se,
$^{78}$Kr, $^{84}$Sr, and $^{92}$Mo can be produced in p-QSE enough to
account for their solar amounts \citep[6th column in
Table~4,][]{Wana2009}. However, these additional sources still cannot
fill the gap for some light p-isotopes such as $^{94}$Mo, $^{96,
  98}$Ru, $^{102}$Pd, $^{106, 108}$Cd, $^{113}$In, $^{115}$Sn, and for
a heavy p-isotope $^{152}$Gd.

Our result in this study is based on a semi-analytic model of
neutrino-driven winds, while the results for the $\gamma$-process, the
$\nu$-process, and the p-QSE listed in Table~4 are all based on
realistic hydrodynamic studies. Nevertheless, we attempt to present a
list of the p-isotopes whose origin can be attributed to the $\nu
p$-process, as follows.  The requisite overproduction factor for a
given nuclide \textit{per supernova event}, which explains its solar
origin, is inferred to be $> 10$ \citep[e.g.,][]{Woos1994}. Assuming
the masses of the total ejecta and of the neutrino-driven ejecta to be
$\sim 10\, M_\odot$ and $\sim 10^{-3}\, M_\odot$
\citep[e.g.,][]{Wana2006}, the overproduction factor per supernova
event is diluted by about 4 orders of magnitude compared to our
result.  We thus apply the condition $f > 10^5$ and $f >
f_\mathrm{max}/10$ to each p-isotope abundance in Figure~6 (the
standard model with $Y_\mathrm{e, 3}$ ranging between 0.5 and 0.7).

The p-isotopes that satisfy the above condition are listed in the last
column of Table~4. According to recent hydrodynamic studies
\citep{Fisc2010, Hued2010}, the maximum $Y_\mathrm{e}$ in the
neutrino-driven outflows is $\sim 0.6$. Therefore, the p-isotopes that
satisfy the above condition only with $Y_\mathrm{e, 3} > 0.6$ are
indicated by ``[yes]''. This implies that the $\nu$p-process in
core-collapse supernovae is the possible astrophysical origin of the
light p-nuclei up to $A = 108$. In principle, however, the $\nu
$p-process can account for the origin of the heavy p-isotopes up to $A
= 152$ as well, if $Y_\mathrm{e, 3} \approx 0.65$ (Figure~6) is
achieved in the neutrino-driven outflows. If this is true, a
reasonable combination of the astrophysical sources considered here
can explain all the origins of the solar p-isotopes. It should be
noted that most of the maximum production factors of these heavy
p-nuclei are $\gtrsim 10^8$. This is three orders of magnitude larger
than the above requisite value ($f = 10^5$). Thus, only $\sim 0.1\%$
of neutrino-driven ejecta with $Y_\mathrm{e, 3} \approx 0.60-0.65$ is
enough to account for the origin of these heavy p-nuclei. Future
multi-dimensional hydrodynamic studies of core-collapse supernovae
with full neutrino transport will be of particular importance if such
a condition is indeed obtained.

A word of caution for the molybdenum isotopes is needed here. The
production factors of $^{92}$Mo and $^{94}$Mo satisfy the above
condition only marginally with $Y_\mathrm{e, 3} = 0.53-0.54$. The
future measurements of the nuclear masses of $^{82}$Zr and $^{83}$Nb
might in part cure this problem as discussed in \S~5.3. This is rather
serious for the origin of $^{94}$Mo that can be produced only by the
$\nu$p-process, while $^{92}$Mo can be explained by the
p-QSE. \citet{Fisk2009} concluded that the ratio $^{92}$Mo/$^{94}$Mo
is about 5 times smaller than the solar value, when applying the
proton separation energy of $^{93}$Rh in \citet{Webe2008}. This might
implies that $^{92}$Mo has another origin, presumably the p-QSE. We
however obtain a reasonable ratio with our standard model (see, e.g.,
the bottom panel of Figure~22) and many other cases (see the
$Y_\mathrm{e, 3} \le 0.55$ range in Figure~6). This is due to the
significant role of $^{92}$Ru$(n, p)^{92}$Tc that competes with
$^{92}$Ru$(p, \gamma)^{93}$Rh in our cases. This is a consequence of
the values of $\Delta_\mathrm{n}$ in the present cases being about a
factor of three higher than those in \citet{Prue2006}. This suggests
that $^{92}$Mo/$^{94}$Mo is highly sensitive to the details of
supernova dynamics.

\section{Summary}

We investigated the effects of uncertainties in supernova dynamics as
well as in nuclear data inputs on the $\nu$p-process in the
neutrino-driven outflows of core-collapse supernovae. The former
includes the wind-termination radius $r_\mathrm{wt}$ (or temperature
$T_\mathrm{wt}$), neutrino luminosity $L_\nu$, neutron-star mass
$M_\mathrm{ns}$, and electron fraction $Y_\mathrm{e, 9}$ (or
$Y_\mathrm{e, 3}$, at $T_9 = 9$ and 3, respectively). The latter
includes the reactions relevant to the breakout from the pp-chain
region ($A < 12$), the $(n, p)$ reactions on heavy nuclei ($Z \ge
56$), and the nuclear masses ($40 \le Z \le 50$) on the $\nu$p-process
pathway. Our result is summarized as follows.

1. Wind termination of the neutrino-driven outflow by colliding with
the preceding supernova ejecta causes a slowdown of the temperature
decrease and thus plays a crucial role on the $\nu$p-process. The
termination within the temperature range of $T_9 = 1.5-3$ (relevant to
the $\nu$p-process) substantially increases the number of neutrons
captured by the seed nuclei ($\Delta_\mathrm{n}$) and thus enhances
efficiency of the p-nuclei production. In the current case, the
efficiency is maximal at $T_\mathrm{wt, 9} = 2.65$ ($r_\mathrm{wt} =
231$~km for $L_\nu = 10^{52}$~erg~s$^{-1}$). This implies that the
early wind phase with the termination radius close to the
proto-neutron star surface is favored for the $\nu$p-process.

2. A lower $L_\nu$ (with the other parameters $T_\mathrm{wt}$,
$M_\mathrm{ns}$, and $Y_\mathrm{e, 9}$ unchanged) leads to more
efficient $\nu$p-processing. This is due to the larger entropy per
nucleon for a lower $L_\nu$, which increases the proton-to-seed ratio
$Y_\mathrm{p}/Y_\mathrm{h}$ and thus $\Delta_\mathrm{n}$. However, the
role of the wind termination is more crucial and thus we presume that
the maximum efficiency is obtained during the early phase with $L_\nu
\sim 10^{52}$~erg~s$^{-1}$.

3. A larger $M_\mathrm{ns}$ (with the other parameters
$T_\mathrm{wt}$, $L_\nu$, and $Y_\mathrm{e, 9}$ unchanged) results in
a larger efficiency of the $\nu$p-process. This is a consequence of
the larger entropy per nucleon and the faster expansion of the
neutrino-driven outflow for a larger $M_\mathrm{ns}$, both of which
help to increase $Y_\mathrm{p}/Y_\mathrm{h}$ and thus
$\Delta_\mathrm{n}$. This implies that a more massive progenitor is
favored for more efficient $\nu$p-processing, if other parameters
remain unchanged. In reality, however, the evolutions of $L_\nu$,
$r_\mathrm{wt}$, and $Y_\mathrm{e}$ will be dependent on the
progenitor mass, making it difficult to draw definitive conclusions.

4. The $\nu$p-process is highly sensitive to the electron fraction
$Y_\mathrm{e, 3}$ that controls $Y_\mathrm{p}/Y_\mathrm{h}$ at the
onset of the $\nu$p-process and thus $\Delta_\mathrm{n}$. An increase
of only $\Delta Y_\mathrm{e, 3} \sim 0.03$ results in $\Delta
A_\mathrm{max} \sim 10$. The models with $Y_\mathrm{e, 3} = 0.52-0.60$
(with the other parameters unchanged) produce sufficient amounts of
the light p-nuclei up to $A = 108$. Furthermore, the models with
$Y_\mathrm{e, 3} = 0.60-0.65$ produce the p-nuclei up to $A =
152$. Note that this is a combined effect of the high $Y_\mathrm{e,
  3}$ and the wind termination at sufficiently high temperature
($T_\mathrm{wt, 9} = 2.19$ in the standard model). Our result shows no
substantial enhancement of the p-nuclei with $A > 152$, since the
nuclear flow reaches the $\beta$-stability line and enters to the
neutron-rich region at $A \sim 130-160$. This is a consequence that a
large $\Delta_\mathrm{n}$ leads to the strong $(n, \gamma)$ flows that
compete with those by $(p, \gamma)$ for $Z > 50$.

5. Variations on the nuclear reactions relevant to the breakout from
the pp-chain region ($A < 12$), namely of triple-$\alpha$,
$^7$Be$(\alpha, \gamma)^{11}$C, and $^{10}$B$(\alpha, p)^{13}$C affect
the $\nu $p-process by changing $Y_\mathrm{p}/Y_\mathrm{h}$ (and
$\Delta_\mathrm{n}$) or producing intermediate-mass nuclei (proton
poison) during $\nu $p-processing. Among these reactions,
triple-$\alpha$ has the largest impact, although the other two show
non-negligible effects, on the production of the p-nuclei at $A \sim
100-110$.

6. Variations on the $(n, p)$ reactions on $^{56}$Ni (seed nuclei),
$^{60}$Zn, and $^{64}$Ge (first $\beta^+$-waiting point on the
classical rp-process) show great impact on efficiency of the $\nu
$p-process for heavy element synthesis. Only a factor of two variation
leads to a factor of 10 or more changes in the production of the
p-nuclei with $A \sim 100-110$ for the first reaction (but somewhat
smaller changes for the latter two reactions). This is a consequence
that these reactions control the strength of the nuclear flow passing
through the $(n, p)$-waiting points ($^{56}$Ni, $^{60}$Zn, and
$^{64}$Ge) on the $\nu$p-process path. We also find that the $N=50$
nucleus $^{96}$Pd plays a role of the ``second seed'' for production
of heavier nuclei.

7. Application of the new experimental masses of \citet[][for $39 \le
Z \le 46$]{Webe2008} exhibits a suppression of the flow $^{87}$Mo$(p,
\gamma)^{88}$Tc ($N=45$), which however do not affect the
nucleosynthetic p-abundances. Our sensitivity tests for unmeasured
nuclear masses indicate that a future measurement of the $^{82}$Zr
mass (and of $^{93}$Nb with a large estimated error) on $N=42$ could
reduce the abundance of $^{84}$Sr by a factor of two.

8. Our result implies that, within possible ranges of uncertainties in
supernova dynamics as well as in nuclear data inputs, the solar
inventory of the light p-nuclei up to $A = 108$ ($^{108}$Cd) can be
attributed to the $\nu$p-process, including the most mysterious ones
$^{92, 94}$Mo and $^{96, 98}$Ru. The molybdenum isotopes are, however,
tend to be underproduced compared to the neighboring p-isotopes. If
highly proton-rich conditions with $Y_\mathrm{e, 3} = 0.60-0.65$ are
realized in neutrino-driven ejecta, the solar origin of the p-nuclei
up to $A = 152$ ($^{152}$Gd) can be explained by the $\nu$p-process.

Our explorations in this study suggest that more refinements both in
supernova conditions and in nuclear data inputs are needed to
elucidate the role of the $\nu$p-process as the astrophysical origin
of the p-nuclei. In particular, multi-dimensional studies of
core-collapse simulations with full neutrino transport, as well as
experiment-based rates of triple-$\alpha$ and the $(n, p)$ reactions
on heavy nuclei will be important in the future works.

\acknowledgements

We are grateful to T. Shima for useful discussion on the
triple-$\alpha$ rate. The project was supported by the Deutsche
Forschungsgemeinschaft through Cluster of Excellence EXC~153 ``Origin
and Structure of the Universe'' (http://www.universe-cluster.de).

\begin{deluxetable}{ccccccccccccccccc}
\tabletypesize{\scriptsize}
\tablecaption{Results for Various Wind Models}
\tablewidth{0pt}
\tablehead{
\colhead{$M_\mathrm{ns}$} &
\colhead{$\log L_\nu$} &
\colhead{$r_\mathrm{wt}$} &
\colhead{$Y_\mathrm{e, 9}$\tablenotemark{a}} &
\colhead{$\dot{M}$} &
\colhead{$S$} &
\colhead{$\tau_1$\tablenotemark{b}} &
\colhead{$\tau_2$\tablenotemark{c}} &
\colhead{$T_\mathrm{wt, 9}$\tablenotemark{d}} &
\colhead{$Y_\mathrm{e, 3}$\tablenotemark{e}} &
\colhead{$Y_\mathrm{p}/Y_\mathrm{h}$\tablenotemark{f}} &
\colhead{$n_{\bar{\nu}_\mathrm{e}}$} &
\colhead{$\Delta_n$} &
\colhead{$\log f_\mathrm{max}$\tablenotemark{g}} &
\colhead{nuc($f_\mathrm{max}$)\tablenotemark{h}} &
\colhead{nuc($A_\mathrm{max}$)\tablenotemark{i}} &
\colhead{Fig.} \\
\colhead{[$M_\odot$]} &
\colhead{[erg s$^{-1}$]} &
\colhead{[100 km]} &
\colhead{} &
\colhead{[$10^{-4} M_\odot$]} &
\colhead{[$k_\mathrm{B}$]} &
\colhead{[ms]} &
\colhead{[ms]} &
\colhead{} &
\colhead{} &
\colhead{} &
\colhead{} &
\colhead{} &
\colhead{} &
\colhead{} &
\colhead{}
}
\startdata
1.4 & 52.0 & 3.00 & 0.600 & 2.70 & 57.0 & 17.5 & 245 & 2.19 & 0.550 & 124 & 0.0834 & 10.3  & 7.16 & $^{96}$Ru & $^{106}$Cd & all \\ 
\\
1.4 & 52.0 & 1.00 & 0.600 & 2.70 & 57.0 & 359 & 1160 & 5.19 & 0.509 & 1.78 & 0.135 & 0.240  & 4.44 & $^{64}$Zn & $^{74}$Se & 2 \\
1.4 & 52.0 & 2.00 & 0.600 & 2.70 & 57.0 & 17.5 & 516 & 2.95 & 0.550 & 124 & 0.138 & 17.1  & 6.27 & $^{78}$Kr & $^{84}$Sr & 2 \\
1.4 & 52.0 & 2.31 & 0.600 & 2.70 & 57.0 & 17.5 & 403 & 2.65 & 0.550 & 124 & 0.114 & 14.1  & 7.67 & $^{106}$Cd & $^{108}$Cd & 2 \\
1.4 & 52.0 & 3.00 & 0.600 & 2.70 & 57.0 & 17.5 & 245 & 2.19 & 0.550 & 124 & 0.0834 & 10.3  & 7.16 & $^{96}$Ru & $^{106}$Cd & 2 \\ 
1.4 & 52.0 & 4.00 & 0.600 & 2.70 & 57.0 & 17.5 & 117 & 1.80 & 0.550 & 124 & 0.0628 & 7.79  & 6.86 & $^{84}$Sr & $^{102}$Pd & 2 \\
1.4 & 52.0 & 5.00 & 0.600 & 2.70 & 57.0 & 17.5 & 44.0 & 1.55 & 0.550 & 124 & 0.0529 & 6.56  & 6.69 & $^{84}$Sr & $^{84}$Sr & 2 \\
1.4 & 52.0 & 10.0 & 0.600 & 2.70 & 57.0 & 17.5 & 30.0 & 1.21 & 0.550 & 124 & 0.0431 & 5.34  & 6.13 & $^{78}$Kr & $^{84}$Sr & 2 \\
1.4 & 52.0 & $\infty$ & 0.600 & 2.70 & 57.0 & 17.5 & 30.0 & ---- & 0.550 & 124 & 0.0323 & 4.01  & 5.79 & $^{78}$Kr & $^{84}$Sr & 2 \\
\\
1.4 & 52.4 & 8.01 & 0.600 & 31.3 & 33.7 & 22.8 & 261 & 2.19 & 0.558 & 42.7 & 0.0611 & 2.61  & 6.00 & $^{78}$Kr & $^{84}$Sr & 3 \\
1.4 & 52.2 & 4.29 & 0.600 & 8.66 & 44.7 & 18.9 & 236 & 2.19 & 0.554 & 78.3 & 0.0720 & 5.64  & 6.83 & $^{84}$Sr & $^{96}$Ru & 3 \\
1.4 & 52.0 & 3.00 & 0.600 & 2.70 & 57.0 & 17.5 & 245 & 2.19 & 0.550 & 124 & 0.0834 & 10.3  & 7.16 & $^{96}$Ru & $^{106}$Cd & 3 \\ 
1.4 & 51.8 & 2.22 & 0.600 & 0.921 & 70.1 & 17.8 & 262 & 2.19 & 0.545 & 166 & 0.0945 & 15.7  & 7.78 & $^{102}$Pd & $^{108}$Cd & 3 \\
1.4 & 51.6 & 1.71 & 0.600 & 0.339 & 83.3 & 19.9 & 301 & 2.19 & 0.540 & 185 & 0.107 & 19.8  & 7.99 & $^{106}$Cd & $^{108}$Cd & 3 \\
1.4 & 51.4 & 1.37 & 0.600 & 0.131 & 96.3 & 24.4 & 371 & 2.19 & 0.535 & 174 & 0.121 & 21.1  & 8.07 & $^{106}$Cd & $^{108}$Cd & 3 \\
\\
1.2 & 52.0 & 3.27 & 0.600 & 3.96 & 46.8 & 18.4 & 241 & 2.19 & 0.553 & 84.4 & 0.0746 & 6.30  & 6.96 & $^{84}$Sr & $^{102}$Pd & 4 \\
1.4 & 52.0 & 3.00 & 0.600 & 2.70 & 57.0 & 17.5 & 245 & 2.19 & 0.550 & 124 & 0.0834 & 10.3  & 7.16 & $^{96}$Ru & $^{106}$Cd & 4 \\ 
1.6 & 52.0 & 2.80 & 0.600 & 1.94 & 68.1 & 16.4 & 244 & 2.19 & 0.547 & 178 & 0.0908 & 16.2  & 7.69 & $^{102}$Pd & $^{108}$Cd & 4 \\
1.8 & 52.0 & 2.62 & 0.600 & 1.46 & 80.0 & 15.4 & 245 & 2.19 & 0.545 & 243 & 0.0980 & 23.8  & 7.91 & $^{106}$Cd & $^{108}$Cd & 4 \\
2.0 & 52.0 & 2.46 & 0.600 & 1.13 & 93.0 & 14.4 & 247 & 2.19 & 0.543 & 335 & 0.104 & 34.8  & 8.12 & $^{108}$Cd & $^{108}$Cd & 4 \\
\\
1.4 & 52.0 & 3.00 & 0.550 & 2.70 & 57.0 & 17.5 & 245 & 2.19 & 0.523 & 42.9 & 0.0834 & 3.58  & 6.25 & $^{78}$Kr & $^{84}$Sr & 5 \\
1.4 & 52.0 & 3.00 & 0.600 & 2.70 & 57.0 & 17.5 & 245 & 2.19 & 0.550 & 124 & 0.0834 & 10.3  & 7.16 & $^{96}$Ru & $^{106}$Cd & 5 \\ 
1.4 & 52.0 & 3.00 & 0.650 & 2.70 & 57.0 & 17.5 & 245 & 2.19 & 0.576 & 245 & 0.0834 & 20.4  & 8.14 & $^{106}$Cd & $^{108}$Cd & 5 \\
1.4 & 52.0 & 3.00 & 0.700 & 2.70 & 57.0 & 17.5 & 245 & 2.19 & 0.603 & 428 & 0.0834 & 35.7  & 8.34 & $^{108}$Cd & $^{120}$Te & 5 \\
1.4 & 52.0 & 3.00 & 0.750 & 2.70 & 57.0 & 17.5 & 245 & 2.19 & 0.629 & 703 & 0.0834 & 58.6  & 8.54 & $^{138}$La & $^{138}$La & 5 \\
1.4 & 52.0 & 3.00 & 0.800 & 2.70 & 57.0 & 17.5 & 245 & 2.19 & 0.655 & 1130 & 0.0834 & 94.2  & 8.37 & $^{138}$La & $^{152}$Gd & 5 
\enddata
\tablenotetext{a}{$Y_\mathrm{e}$ at $T_9 = 9$.}
\tablenotetext{b}{time elapsed from $T_9 = 6$ to $T_9 = 3$.}
\tablenotetext{c}{time elapsed from $T_9 = 3$ to $T_9 = 1.5$.}
\tablenotetext{d}{temperature (in units of 10$^9$ K) just after the wind-termination.}
\tablenotetext{e}{$Y_\mathrm{e}$ at $T_9 = 3$.}
\tablenotetext{f}{proton-to-seed ratio at $T_9 = 3$.}
\tablenotetext{g}{maximum production factor.}
\tablenotetext{h}{nuclide at $f = f_\mathrm{max}$.}
\tablenotetext{i}{nuclide at the largest $A$ with $f > f_\mathrm{max}/10$.}
\end{deluxetable}

\begin{deluxetable}{ccccc}
\tablecaption{Rates and decay timescales for selected reactions}
\tablewidth{0pt}
\tablehead{
\colhead{Species} &
\colhead{$\lambda_\mathrm{2.5}$\tablenotemark{a}} &
\colhead{$\tau_\mathrm{2.5}$\tablenotemark{a}} &
\colhead{$\lambda_\mathrm{2.0}$\tablenotemark{b}} &
\colhead{$\tau_\mathrm{2.0}$\tablenotemark{b}} \\
\colhead{} &
\colhead{[mol$^{-1}$ cm$^3$ s$^{-1}$]} &
\colhead{[ms]} &
\colhead{[mol$^{-1}$ cm$^3$ s$^{-1}$]} &
\colhead{[ms]} 
}
\startdata
$3\alpha$ & $3.07\times 10^{-10}$ & $8.32\times 10^4$ & $3.86\times
 10^{-10}$ & $2.96\times 10^5$  \\
$^{7}$Be$(\alpha, \gamma)$ & 4.48 & $1.45\times 10^{-2}$ & 1.29 &
 0.107 \\
$^{7}$Be$(\alpha, p)$ & $4.22\times 10^4$ & $1.55\times 10^{-6}$ &
 $5.89\times 10^3$ & $2.34\times 10^{-5}$ \\
$^{10}$B$(\alpha, p)$ & $5.76\times 10^5$ & $1.13\times 10^{-7}$ &
 $1.36\times 10^5$ & $1.01\times 10^{-6}$ \\
$^{10}$B$(p, \alpha)$ & $1.14\times 10^7$ & $1.25\times 10^{-8}$ &
 $6.04\times 10^6$ & $5.59\times 10^{-8}$ \\
$^{11}$C$(\alpha, p)$ & $6.98\times 10^4$ & $9.34\times 10^{-7}$ &
 $1.61\times 10^4$ & $8.58\times 10^{-6}$ \\
$^{12}$C$(p, \gamma)$ & $4.69\times 10^2$ & $3.05\times 10^{-4}$ &
 $5.11\times 10^2$ & $6.59\times 10^{-4}$ \\
$^{13}$C$(\alpha, n)$ & $2.01\times 10^4$ & $3.24\times 10^{-6}$ &
 $1.27\times 10^4$ & $1.09\times 10^{-5}$ \\
$^{56}$Ni$(n, p)$ & $1.61\times10^8$ & 11.3 & $1.28\times10^8$ & 454 \\
$^{60}$Zn$(n, p)$ & $7.40\times10^8$ & 2.46 & $6.82\times10^8$ & 85.5 \\
$^{64}$Ge$(n, p)$ & $6.85\times10^8$ & 2.65 & $6.15\times10^8$ & 94.8 \\
$^{68}$Se$(n, p)$ & $9.38\times10^8$ & 1.94 & $8.12\times10^8$ & 71.8 \\
$^{72}$Kr$(n, p)$ & $1.38\times10^9$ & 1.32 & $1.24\times10^9$ & 46.9 \\
$^{76}$Sr$(n, p)$ & $1.81\times10^9$ & 1.00 & $1.69\times10^9$ & 34.6 \\
$^{80}$Zr$(n, p)$ & $1.99\times10^9$ & 0.913 & $1.92\times10^9$ & 30.4 \\
$^{84}$Mo$(n, p)$ & $1.44\times10^9$ & 1.26 & $1.31\times10^9$ & 44.5 \\
$^{96}$Pd$(n, p)$ & $2.06\times10^8$ & 8.84 & $1.66\times10^8$ & 351 
\enddata
\tablenotetext{a}{Rates and decay timescales at $T_9 = 2.5$ ($\rho =
 7.20\times 10^4$~g~cm$^{-3}$, $X_\mathrm{n} = 7.64\times 10^{-12}$, $X_\mathrm{p} = 0.0970$, $X_\alpha = 0.852$).}
\tablenotetext{b}{Rates and decay timescales at $T_9 = 2.0$ ($\rho =
 3.46\times 10^4$~g~cm$^{-3}$, $X_\mathrm{n} = 4.96 \times 10^{-13}$, $X_\mathrm{p} = 0.0858$, $X_\alpha = 0.838$).}
\end{deluxetable}

\begin{deluxetable}{cccccccc}
\tablecaption{Results for the Changes of Reaction Rates}
\tablewidth{0pt}
\tablehead{
\colhead{reaction} &
\colhead{factor} &
\colhead{$Y_\mathrm{p}/Y_\mathrm{h}$\tablenotemark{a}} &
\colhead{$\Delta_n$\tablenotemark{b}} &
\colhead{$\log f_\mathrm{max}$\tablenotemark{c}} &
\colhead{nuc($f$)\tablenotemark{d}} &
\colhead{nuc($A$)\tablenotemark{e}} &
\colhead{Fig.}
}
\startdata
standard & 1.00 & 124 & 10.3 & 7.16 & $^{96}$Ru & $^{106}$Cd & all \\
3$\alpha$ & 1.00\tablenotemark{f} & 25.6 & 2.14 & 6.47 & $^{78}$Kr & $^{84}$Sr & 12 \\
3$\alpha$ & 2.00 & 73.5 & 6.13 & 6.93 & $^{84}$Sr & $^{102}$Pd & 12 \\
3$\alpha$ & 10.0 & 25.2 & 2.10 & 6.15 & $^{78}$Kr & $^{84}$Sr & 12 \\
3$\alpha$ & 1/2.00 & 204 & 17.0 & 7.67 & $^{102}$Pd & $^{108}$Cd & 12 \\
3$\alpha$ & 1/10.0 & 482 & 40.2 & 8.04 & $^{108}$Cd & $^{108}$Cd & 12 \\
3$\alpha$ & 1/100 & 719 & 60.0 & 8.02 & $^{108}$Cd & $^{120}$Te & 12 \\
$^7$Be($\alpha$, $\gamma$) & 2.00 & 124 & 10.3 & 7.11 & $^{96}$Ru & $^{106}$Cd & 13 \\
$^7$Be($\alpha$, $\gamma$) & 10.0 & 122 & 10.2 & 6.98 & $^{96}$Ru & $^{106}$Cd & 13 \\
$^7$Be($\alpha$, $\gamma$) & 100 & 117 & 9.76 & 6.89 & $^{84}$Sr & $^{106}$Cd & 13 \\
$^7$Be($\alpha$, $\gamma$) & 1/2.00 & 124 & 10.3 & 7.19 & $^{96}$Ru & $^{106}$Cd & 13 \\
$^7$Be($\alpha$, $\gamma$) & 1/10.0 & 124 & 10.3 & 7.24 & $^{102}$Pd & $^{106}$Cd & 13 \\
$^{10}$B($\alpha$, $p$) & 2.00 & 119 & 9.92 & 7.09 & $^{96}$Ru & $^{106}$Cd & 14 \\
$^{10}$B($\alpha$, $p$) & 10.0 & 112 & 9.34 & 6.96 & $^{96}$Ru & $^{106}$Cd & 14 \\
$^{10}$B($\alpha$, $p$) & 1/2.00 & 129 & 10.8 & 7.23 & $^{102}$Pd & $^{106}$Cd & 14 \\
$^{10}$B($\alpha$, $p$) & 1/10.0 & 135 & 11.3 & 7.35 & $^{102}$Pd & $^{106}$Cd & 14 \\
$^{56}$Ni($n$, $p$) & 2.00 & 124 & 10.3 & 7.01 & $^{96}$Ru & $^{106}$Cd & 16 \\
$^{56}$Ni($n$, $p$) & 10.0 & 124 & 10.3 & 7.02 & $^{84}$Sr & $^{102}$Pd & 16 \\
$^{56}$Ni($n$, $p$) & 1/2.00 & 124 & 10.3 & 7.45 & $^{102}$Pd & $^{106}$Cd & 16 \\
$^{56}$Ni($n$, $p$) & 1/10.0 & 124 & 10.3 & 7.92 & $^{106}$Cd & $^{108}$Cd & 16 \\
$^{60}$Zn($n$, $p$) & 2.00 & 124 & 10.3 & 7.15 & $^{96}$Ru & $^{106}$Cd & 19 \\
$^{60}$Zn($n$, $p$) & 10.0 & 124 & 10.3 & 7.15 & $^{96}$Ru & $^{106}$Cd & 19 \\
$^{60}$Zn($n$, $p$) & 1/2.00 & 124 & 10.3 & 7.22 & $^{102}$Pd & $^{106}$Cd & 19 \\
$^{60}$Zn($n$, $p$) & 1/10.0 & 124 & 10.3 & 7.51 & $^{102}$Pd & $^{108}$Cd & 19 \\
$^{64}$Ge($n$, $p$) & 2.00 & 124 & 10.3 & 7.16 & $^{96}$Ru & $^{106}$Cd & 20 \\
$^{64}$Ge($n$, $p$) & 10.0 & 124 & 10.3 & 7.16 & $^{96}$Ru & $^{106}$Cd & 20 \\
$^{64}$Ge($n$, $p$) & 1/2.00 & 124 & 10.3 & 7.19 & $^{102}$Pd & $^{106}$Cd & 20 \\
$^{64}$Ge($n$, $p$) & 1/10.0 & 124 & 10.3 & 7.41 & $^{102}$Pd & $^{108}$Cd & 20
\enddata
\tablenotetext{a}{proton-to-seed ratio at $T_9 = 3$.}
\tablenotetext{b}{$\Delta_n$ at $T_9 = 3$.}
\tablenotetext{c}{maximum production factor.}
\tablenotetext{d}{nucleus at $f = f_\mathrm{max}$.}
\tablenotetext{e}{nucleus at the largest $A$ with $f > f_\mathrm{max}/10$.}
\tablenotetext{f}{triple-$\alpha$ rate from \citet{Angu1999}.}
\end{deluxetable}

\begin{deluxetable}{cllcccc}
\tablecaption{p-Nuclei Abundances and Their Possible Sources}
\tablewidth{0pt}
\tablehead{
\colhead{Species} &
\colhead{Abundance\tablenotemark{a}} &
\colhead{fraction\tablenotemark{b} [\%]} &
\colhead{$\gamma$-process\tablenotemark{c}} &
\colhead{$\nu$-process\tablenotemark{d}} &
\colhead{p-QSE\tablenotemark{e}} &
\colhead{$\nu$p-process\tablenotemark{f}}
}
\startdata
$^{ 74}$Se & 0.58       & 0.889 & yes & no  & yes & yes \\
$^{ 78}$Kr & 0.20       & 0.362 & yes & no  & yes & yes \\
$^{ 84}$Sr & 0.13124    & 0.5551 & yes & no  & yes & yes \\
$^{ 92}$Mo & 0.386      & 14.8362 & no  & no  & yes & yes \\
$^{ 94}$Mo & 0.241      & 9.2466 & no  & no  & no & yes \\
$^{ 96}$Ru & 0.1053     & 5.542 & no  & no  & no & yes \\
$^{ 98}$Ru & 0.0355     & 1.8688 & no  & no  & no & yes \\
$^{102}$Pd & 0.0146     & 1.02 & no  & no  & no & yes \\
$^{106}$Cd & 0.01980    & 1.25 & no  & no  & no & yes \\
$^{108}$Cd & 0.01410    & 0.89 & no  & no  & no & yes \\
$^{113}$In & 0.0078     & 4.288 & no  & no  & no & [yes] \\
$^{112}$Sn & 0.03625    & 0.971 & [yes] & no  & no & [yes] \\
$^{114}$Sn & 0.02460    & 0.659 & [yes] & no  & no & [yes] \\
$^{115}$Sn & 0.01265    & 0.339 & no  & no  & no & [yes] \\
$^{120}$Te & 0.0046     & 0.096 & [yes] & no  & no & [yes] \\
$^{124}$Xe & 0.00694    & 0.129 & [yes] & no  & no & [yes] \\
$^{126}$Xe & 0.00602    & 0.112 & yes & no  & no & [yes] \\
$^{130}$Ba & 0.00460    & 0.1058 & yes & no  & no & [yes] \\
$^{132}$Ba & 0.00440    & 0.1012 & yes & no  & no & [yes] \\
$^{138}$La & 0.000397   & 0.09017 & no  & yes  & no & [yes] \\
$^{136}$Ce & 0.00217    & 0.186 & yes & no  & no & [yes] \\
$^{138}$Ce & 0.00293    & 0.251 & yes & no  & no & [yes] \\
$^{144}$Sm & 0.00781    & 3.0734 & yes & no  & no & [yes] \\
$^{152}$Gd & 0.00067    & 0.2029 & no  & no  & no & [yes] \\
$^{156}$Dy & 0.000216   & 0.056 & [yes] & no  & no & no \\
$^{158}$Dy & 0.000371   & 0.096 & [yes] & no  & no & no \\
$^{162}$Er & 0.000350   & 0.137 & [yes] & no  & no & no \\
$^{164}$Er & 0.004109   & 1.609 & [yes] & no  & no & no \\
$^{168}$Yb & 0.000323   & 0.13 & yes & no  & no & no \\
$^{174}$Hf & 0.000275   & 0.1620 & yes & no  & no & no \\
$^{180}$Ta & 0.00000258 & 0.0123 & yes & yes  & no & no \\
$^{180}$W  & 0.000153   & 0.1198 & yes & no  & no & no \\
$^{184}$Os & 0.000133   & 0.0198 & yes & no  & no & no \\
$^{190}$Pt & 0.000185   & 0.013634 & yes & no  & no & no \\
$^{196}$Hg & 0.00063    & 0.15344 & yes & no  & no & no 
\enddata
\tablenotetext{a}{\citet{Lodd2003}; Si $=10^6$}
\tablenotetext{b}{\citet{Lodd2003}; relative to its elemental
  abundance} \tablenotetext{c}{\citet{Raye1995} (nuclei indecated by
  ``[yes]'' are those underproduced in \citet{Raus2002})}
\tablenotetext{d}{\citet{Woos1990}}
\tablenotetext{e}{\citet{Wana2009}} \tablenotetext{f}{This work
  (nuclei indicated by ``[yes]'' are produced only with
  $Y_\mathrm{e,3} > 0.6$)}
\end{deluxetable}

\end{document}